\documentclass{cmslatex}
\usepackage[paperwidth=7in, paperheight=10in, margin=.875in]{geometry}
 \usepackage[backref,colorlinks,linkcolor=red,anchorcolor=green,citecolor=blue]{hyperref}
\usepackage{amsfonts,amssymb}
\usepackage{amsmath}
\usepackage{graphicx}
\usepackage{cite}
\usepackage{enumerate}
\usepackage{graphicx,graphics,epsfig,subfigure}
\usepackage{graphicx}
\usepackage{booktabs}
\usepackage{threeparttable}
\usepackage{mathrsfs}
\usepackage{multirow}
\usepackage{appendix}
\usepackage{xcolor,color,soul,array,cite,float,version,times}

\renewcommand{\theequation}{\arabic{section}.\arabic{equation}}

\newcommand{\ben}{\begin{eqnarray}}
\newcommand{\een}{\end{eqnarray}}
\newcommand{\bea}{\begin{array}}
	\newcommand{\eea}{\end{array}}
\newcommand{\bes}{\begin{subequations}}
	\newcommand{\ees}{\end{subequations}}
\newcommand{\bef}{\begin{figure}[H]}
	\newcommand{\eef}{\end{figure}}
\newcommand{\bet}{\begin{tikzpicture}}
\newcommand{\eet}{\end{tikzpicture}}
\newcommand{\beq}{\begin{equation}}
\newcommand{\eeq}{\end{equation}}
\def\bena#1\eena{\begin{eqnarray}\begin{array}{l}#1\end{array}\end{eqnarray}}
\def\besl#1\eesl{\begin{subequations}\begin{align}#1\end{align}\end{subequations}}

\newcommand{\parl}[2]{\ensuremath{\frac{\partial #1}{\partial #2}}}

\def\inc(#1){\includegraphics[height=3 cm]{pics/#1}}

\def\bR{\mathbf{R}}

\def\bv{\mathbf{v}}

\def\bn{\mathbf{n}}
\def\b0{\mathbf{0}}

\def\bx{\mathbf{x}}
\def\by{\mathbf{y}}

\def\bC{\mathbf{C}}

\def\bI{\mathbf{I}}

\def\bM{\mathbf{M}}

\def\bA{\mathbf{A}}

   \allowdisplaybreaks
\begin{document}
 \title{Thermodynamically Consistent Dynamic Boundary Conditions of  Phase Field  Models\thanks{Received date: April 12, 2022 and accepted date (in revised version): August 15, 2022.}}


          \author{Xiaobo Jing\thanks{Beijing Computational Science Research Center, Beijing 100193, P. R. China and Department of Mathematics,
          		University of South Carolina, Columbia, SC 29028, USA, (jingxb@csrc.ac.cn).}
          \and Qi Wang \thanks{Department of Mathematics,
          	University of South Carolina, Columbia, SC 29028, USA, (qwang@math.sc.edu).}}

         \pagestyle{myheadings} \markboth{Thermodynamically Consistent Dynamic Boundary Conditions of  Phase Field  Models}{Xiaobo Jing and Qi Wang} \maketitle

          \begin{abstract}
             We present a general, constructive  method to derive thermodynamically consistent  models and   consistent dynamic boundary conditions hierarchically following the generalized Onsager principle. The method consists of two steps in tandem: the dynamical equation  is determined by the generalized Onsager principle in the bulk firstly,  and then the surface chemical potential and the  thermodynamically consistent boundary conditions are  formulated subsequently by applying the generalized Onsager principle at the boundary. The application strategy of the generalized Onsager principle in two-step yields thermodynamically consistent models together with the consistent boundary conditions that warrant a non-negative entropy production rate (or equivalently non-positive  energy dissipation rate in isothermal cases)  in the bulk as well as at the boundary.
             We illustrate the method using phase field models of binary materials elaborate on two sets of thermodynamically consistent dynamic boundary conditions. These two types of boundary conditions differ in how the across boundary mass flux participates in boundary surface dynamics. We then show that many existing thermodynamically consistent, binary phase field models together with their dynamic or static boundary conditions are derivable from this method.
             As an illustration, we show numerically how dynamic boundary conditions affect crystal growth in the bulk using a binary phase field model.
          \end{abstract}
\begin{keywords} Thermodynamically consistent model ;phase field model; dynamic boundary conditions; binary materials; energy dissipation.
\end{keywords}

\begin{AMS} 35G30; 35G60; 35Q79; 35Q82
\end{AMS}
\section{Introduction} \label{Intro}

\noindent \indent Thermodynamically consistent models refer to the models that are derived from thermodynamical laws and principles. In particular, they obey the second law of thermodynamics, i.e., the entropy production rate  is non-negative or equivalently the energy dissipation rate is non-positive in isothermal cases. The generalized Onsager principle is a protocol in which the Onsager linear response theory combined with the equilibrium maximum entropy principle is used to derive thermodynamically consistent models \cite{onsager1931reciprocal1, onsager1931reciprocal2, yang2016hydrodynamic,Wang2020}. The generalized Onsager  principle warrants the second law of thermodynamics in the form of Clausius-Duhem inequality and has proven to be an effective modeling tool for developing thermodynamical models at various time and length scales \cite{yang2016hydrodynamic, li2019energy,zhao2018thermodynamically,zhao2018general,sun2020structure}. In the past, the Onsager principle and the equivalent thermodynamical second law has been primarily used to derive dynamic equations in the bulk while boundary effects are largely ignored by assuming adiabatic and static boundary conditions or periodic boundary conditions. In this study, we present a general, hierarchical method to derive thermodynamically consistent models with consistent dynamic boundary conditions for material systems using the generalized Onsager principle in not only the bulk but also the boundaries, where the dynamic boundary conditions reduce to  static ones in the limit. We illustrate the idea by deriving  thermodynamically consistent phase field models and consistent boundary conditions for binary materials in a piecewise smooth domain owing to their abundance in the literature.

Phase field modeling has emerged as one of the powerful and versatile modeling paradigm in dealing with multiphase materials in domains with  complex interface geometries and complex interfacial phenomena between distinct immiscible phases \cite{xing2009topology,steinbach1996phase,singer2008phase}. It is especially useful and effective when handling dynamic phase boundaries in multiphase materials involving topological changes compared to other methods such as front tracking methods, level-set methods, and volume of fluid methods etc\cite{steinbach2009phase, chen2002phase,singer2008phase}. By design, it is for diffuse interfaces with certain interface thickness in which complex interfacial dynamics prevails.   A quality phase field model should be able to capture   well-known sharp interface conditions (e.g Gibbs-Thomson condition) in the vanishing thickness of the interface \cite{galenko2019rapid,salhoumi2016gibbs}. This requires one to be mindful when deriving the free energy of the system in the phase field model so that thermodynamical laws and principles are followed faithfully and properly.  Notice that the advantage of the phase field model in dealing with diffuse interfaces may also limit its applicability to resolve sharp
interfaces \cite{elder2001sharp}. In life science, materials science and engineering, there are multiphase material systems with diffuse interfaces, which have kept multiphase field models popular and practical \cite{nestler2011phase, steinbach2009phase, kim2012phase, provatas2011phase}.

There are quite a  number of phase field models in the literature today. However, not all  are thermodynamically consistent.   In this study, we focus on the derivation of thermodynamically consistent phase field models, which include the derivation of the transport equation in the bulk as well as the consistent dynamic boundary conditions in  fixed boundaries. We stress that it is  important to study multiphase material systems using a thermodynamically "correct" model that not only gives one a comprehensive description of the correct physics, but also gives one a well-posed mathematical system to analyze and compute. Speaking of a thermodynamically "correct" model, we insist that the model must be at least thermodynamically consistent. This humble criterion would perhaps disqualify a host of existing phase field models.
In addition, we notice that most of the studies on phase field models are concentrated on equations in the bulk with static or periodic boundary conditions at fixed boundaries, where the boundary contributions to thermodynamical consistency are trivialized.

Given the recent technological advances in materials science and engineering, boundaries of a material confining device can no longer be treated as passive. They can be made with distinctive properties to interact or even control the  material within the device \cite{yoon2011topology,gavrilyuk2019dynamic,rakita2019defects}. For instance, the newly discovered boundary effect to the existence of blue phases in cholesteric liquid crystals in microscales across a quite large temperature range is one of the prominent examples \cite{martinez2015blue,chen2017large,rahman2015blue}.  This requires one to derive a model for the material system to take into account the potential dynamic contribution from the boundary. There have been a surge in activities of this direction on phase field models recently \cite{colli2019coupled,fukao2019separation,gal2008asymptotic,garcke2020weak,colli2015cahn,fukao2017structure,okumura2020structure}.
Here, we  briefly review  some existing thermodynamically consistent phase field models with various static and dynamic  boundary conditions.

When one studies dynamics of a phase field model in a fixed domain with adiabatic boundary conditions, no-flux boundary conditions are normally adopted as sufficient adiabatic conditions which contribute to a zero energy flux across the boundary. The most commonly studied phase field models are the Allen-Cahn and the Cahn-Hilliard model, respectively \cite{allen1979microscopic,cahn1958free}.
We consider a binary material system in a domain $\Omega$ with the free energy  given by
\bena
E=\int_\Omega [\frac{\epsilon}{2}|\nabla \phi|^2+\frac{1}{\epsilon}f(\phi)] d\bx, \label{1.1}
\eena
where $\phi$ is a mass or volume fraction of one material's component, boundary $\partial \Omega$ is piecewise smooth, $\epsilon$ is the strength of the conformational entropy and $\frac{1}{\epsilon}f(\phi)$ is the bulk free energy density. For simplicity, we refer to $\phi$ as the mass fraction throughout the paper.


The Allen-Cahn equation \cite{allen1979microscopic} for  dynamics of $\phi$ is given by
\ben
\phi_t= -M_b^{(1)}  \mu, \quad \mu=\frac{\delta E}{\delta \phi},\label{1.2}
\een
where $M_b^{(1)}$ is the positive semi-definite  mobility operator and $\mu$ is known as the chemical potential.
For adiabatic boundaries, one uses the following homogeneous Neumann boundary condition (HNBC) to ensure that the energy dissipation of the system is dictated exclusively by bulk energy dissipation
\bena
\bn \cdot \nabla \phi=\nabla_\bn \phi=0. \label{1.3}
\eena
The energy dissipation rate in the model is given by the bulk integral without any boundary contributions
\bena
\frac{d}{dt}E=-\int_\Omega \mu M_b^{(1)}  \mu d\bx. \label{1.4}
\eena
Notice that this model doesn't conserve mass.  To conserve mass, one uses another one known as the Cahn-Hilliard model with homogeneous Neumann boundary condition (HNBC-CH model).


The Cahn-Hilliard equation \cite{cahn1958free} for dynamics of $\phi$ is given by
\ben
\phi_t=\nabla \cdot \bM_b^{(2)}  \cdot \nabla \mu, \quad \mu=\frac{\delta E}{\delta \phi}, \label{1.5}
\een
where $ \bM_b^{(2)} $ is the positive semi-definite mobility  coefficient matrix.
The following homogeneous Neumann boundary conditions  ensure the mass conservation and energy dissipation for the model simultaneously
\bena
\bn\cdot  \bM_b^{(2)}  \cdot \nabla \mu=0, \quad
\bn \cdot \nabla \phi=0, \label{1.6}
\eena
where  the first equation is called the no mass flux condition, resulted from the variation in the energy dissipation rate, and the second  condition  ensures that no boundary energy fluxes result from the conformational entropy. Note that all these are static boundary conditions so that there is no boundary dynamics in this model. This is by far the most widely studied phase field model in the literature \cite{cherfils2011cahn,rybka1999convergence,jing2019second}. Besides the Cahn-Hilliard model, non-local constraints can be added to the Allen-Cahn model to enforce the mass conservation to yield the Allen-Cahn model with nonlocal constraints \cite{jing2019second,jing2020linear}.

If there are material and/or energy exchange across the boundary or there exists dynamics on the boundary, dynamics of materials in the bulk can be affected. Next, we list several binary phase field  models with dynamic boundary conditions studied recently. The free energy of the  binary material system of these models consists of two parts: the bulk free energy $E_b$ and the surface free energy $E_s$, respectively,
\bena
E=E_b+E_s,\quad
E_b=\int_\Omega \frac{\epsilon}{2}|\nabla \phi|^2+\frac{1}{\epsilon}f(\phi)d\bx,\quad
E_s=\int_{\partial \Omega} \frac{\delta}{2}|\nabla_s \phi|^2+\frac{1}{\delta}g(\phi) ds, \label{1.7}
\eena
where $\nabla_s=\nabla- (\bn \cdot \nabla)\bn=(\bI-\bn\bn) \nabla$ is the surface gradient operator\cite{dziuk2013finite,brenner2013interfacial}, $\delta$ is the strength of the conformational entropy at the boundary and $g(\phi)$ is the surface energy density at the boundary\cite{fischer1997novel, fischer1998time,kenzler2001phase}.


Gal et al derived a set of  dynamic boundary conditions for the Cahn-Hilliard model   in \cite{gal2006cahn} (Gal model),
\bena
\parl{\phi}{t}=M_b^{(2)} \Delta \mu, \quad \mu=-\epsilon \Delta \phi+\frac{1}{\epsilon}f'(\phi), \quad \bx \in \Omega, \\
\parl{\phi}{t}=-\mu_s-\beta M_b^{(2)}  \nabla_\bn \mu, \quad  \bx \in \partial \Omega,\\ \mu_s=-\delta\Delta_s \phi+\frac{1}{\delta}g'(\phi)+\epsilon \nabla_\bn \phi, \quad
\mu=\beta \mu_s, \quad \bx \in \partial \Omega,\\
\phi(\bx, 0)=\phi_0(\bx), \quad \bx \in \Omega\cup\partial \Omega, \label{1.8}
\eena
where $\nabla_\bn=\bn \cdot \nabla,$ and  $\Delta_s$ is the Laplace-Beltrami operator \cite{dziuk2013finite,grigoryan2009heat}.
In this model, the mobility operator is given by $M_b^{(2)} \Delta$ with a constant mobility coefficient $M_b^{(2)}$ and
the energy dissipation rate is given by
\ben
\frac{d}{dt}E=-M_b^{(2)} \int_\Omega |\nabla \mu|^2 d\bx-\int_{\partial \Omega} |\mu_s|^2 ds.
\een
At the boundary, the chemical potential from the bulk is stipulated to be proportional to the chemical potential at the surface with proportionality parameter $\beta$; the effective chemical potential at the surface includes a flux contribution from the bulk; the dynamic equation of the mass fraction on the surface
follows Allen-Cahn dynamics with an additional flux from the bulk so that it does not conserve any quantities as seen in other models below.  Neither mass of the bulk nor mass of the boundary is conserved in this model. If we choose $g=\delta =0$ and let $\beta \to \infty$, the Gal model reduces to the HNBC-CH model.


In 2011, Goldstein et al \cite{goldstein2011cahn} modified the boundary transport equation of the mass fraction in the Cahn-Hilliard model to give the following governing system of equations (GMS model)
\bena
\parl{\phi}{t}=M_b^{(2)} \Delta \mu, \quad \mu=-\epsilon \Delta \phi+\frac{1}{\epsilon}f'(\phi), \quad \bx \in \Omega, \\
\parl{\phi}{t}=M_s^{(2)} \Delta_s \mu_s-\beta M_b^{(2)}  \nabla_\bn \mu, \quad  \bx \in \partial \Omega,\\
 \mu_s=-\delta\Delta_s \phi+\frac{1}{\delta}g'(\phi)+\epsilon \nabla_\bn \phi, \quad 
 \mu=\beta \mu_s, \quad  \bx \in \partial \Omega,\\
\phi(\bx,0)=\phi_0(\bx), \quad \bx \in \Omega\cup\partial \Omega, \label{1.9}
\eena
where $M_s^{(2)}$ is the mobility coefficient for the transport equation at the boundary. This model differs from the above model in the transport equation of $\phi$ at the boundary, where an Allen-Cahn equation is replaced by a surface Cahn-Hilliard equation. This modification results in the following
mass equality
\bena
\beta \int_\Omega \phi(t)d\bx+\int_{\partial \Omega}\phi(t)ds=\beta \int_\Omega \phi(0)d\bx+\int_{\partial \Omega}\phi(0)ds,
\label{mass-c0}
\eena
where  $\beta$ can be viewed as a weight of the bulk mass compared to the surface mass.

The energy dissipation rate equation is given by
\bena
\frac{d}{dt}E=-M_b^{(2)} \int_\Omega |\nabla \mu|^2 d\bx-M_s^{(2)} \int_{\partial \Omega} |\nabla_s \mu_s|^2 ds.\label{edisp-2}
\eena
Note that this model differs from the Gal model in one dynamic boundary condition so that they yield different dynamics at the boundary and thereby different energy dissipation rates.


By setting $\nabla_\bn \mu=0$ instead of enforcing $\beta \mu_s=\mu$ at the boundary, Liu and Wu derived another set of boundary conditions for the Cahn-Hilliard model with dynamic boundary conditions (LW model)\cite{liu2019energetic} :
\bena
\parl{\phi}{t}=M_b^{(2)} \Delta \mu, \quad \mu=-\epsilon \Delta \phi+\frac{1}{\epsilon}f'(\phi), \quad \bx \in \Omega, \\
\parl{\phi}{t}=M_s^{(2)} \Delta_s \mu_s, \quad \mu_s=-\delta\Delta_s \phi+\frac{1}{\delta}g'(\phi)+\epsilon \nabla_\bn \phi, \quad \nabla_\bn \mu=0,  \quad \bx \in \partial \Omega,\\
\phi(\bx, 0)=\phi_0(\bx), \quad \bx \in \Omega\cup\partial \Omega. \label{1.13}
\eena
In this model, mass conservation laws in the bulk and in the boundary are held respectively,
\bena
\int_\Omega \phi(\bx, t)d\bx=\int_\Omega \phi(\bx, 0)d\bx,\quad
\int_{\partial \Omega}\phi(\bx, t)ds=\int_{\partial \Omega}\phi(\bx, 0)ds.\label{mass-c1}
\eena

The energy dissipation rate  is analogous to that in the GMS model, given by \eqref{edisp-2}.
This model dictates that there is no material loss through the boundary and the mass in the bulk and on the boundary are conserved, respectively, shown in \eqref{mass-c1}. Although energy dissipation rates in these two models are identical, their numerical values may be different because of the difference in the dynamic boundary conditions.


In 2019, Knopf and Lam presented yet another set of  dynamic boundary conditions for the Cahn-Hilliard equation by extending LW model at the boundary  (KL model) \cite{knopf2019convergence}. The governing equation system and the boundary conditions are given by
\bena
\parl{\phi}{t}=M_b^{(2)}  \Delta \mu, \quad \mu=-\epsilon \Delta \phi+\frac{1}{\epsilon}f'(\phi), \quad \bx \in \Omega, \\
\parl{\psi}{t}=M_s^{(2)} \Delta_s \mu_s, \quad \mu_s=-\delta \Delta_s \psi+\frac{1}{\delta}g'(\phi)+\epsilon H'(\psi) \nabla_\bn \phi, \quad  \bx \in \partial \Omega,\\
\epsilon K\nabla_\bn \phi=H(\psi)-\phi, \quad \nabla_\bn \mu=0, \quad \bx \in \partial \Omega,\\
\phi(\bx,0)=\phi_0(\bx), \quad \bx \in \Omega, \quad \psi(\bx,0)=\psi_0(\bx), \quad \bx\in \partial \Omega, \label{1.15}
\eena
where $H(\psi)$ is a prescribed function of $\psi$. The mass in the bulk and on the boundary are conserved respectively.

Here, a new function $H(\psi)$ is introduced into the chemical potential at the boundary surface.
The modified free energy is given by
\bena
E=\int_\Omega \frac{\epsilon}{2}|\nabla \phi|^2+\frac{1}{\epsilon}f(\phi) d\bx+\int_{\partial \Omega}[ \frac{\delta}{2}|\nabla_s \psi|^2+\frac{1}{\delta}g(\psi)] ds+\int_{\partial \Omega}\frac{(H(\psi)-\phi)^2}{2K}ds, \label{1.16}
\eena
If $K \to 0$ and $H(\psi)=\psi$, the KL model  reduces to the LW model. The last term in the free energy is a term  penalizing  the difference between  $H(\psi)$ and $\phi$ when $K \to 0$.
Recently, Knopf and Signori derived nonlocal models with dynamic boundary conditions and analyzed their well-posedness \cite{knopf2021nonlocal}, extending their work in this direction using distinct variables for the bulk and surface respectively.


In 2019, Knopf et al. derived a set of boundary conditions for the Cahn-Hilliard model (KLLM model) \cite{knopf2020phase}  as follows
\bena
\parl{\phi}{t}=M_b^{(2)} \Delta \mu, \quad \mu=-\epsilon \Delta \phi+\frac{1}{\epsilon}f'(\phi), \quad \bx \in \Omega, \\
\parl{\phi}{t}=M_s^{(2)} \Delta_s \mu_s-\beta M \nabla_\bn \mu, \quad \mu_s=-\delta \Delta_s \phi+\frac{1}{\delta}g'(\phi)+\epsilon \nabla_\bn \phi, \quad \bx \in \partial \Omega,\\
 \alpha \nabla_\bn \mu=\beta \mu_s-\mu, \quad \bx \in \partial \Omega,\\
\phi(\bx,0)=\phi_0(\bx), \quad \bx \in \Omega\cup\partial \Omega, \label{KLLM}
\eena
where $\alpha$ is a relaxation length parameter. Instead of forcing $\mu=\beta \mu_s$ or $\nabla_\bn \mu=0$ at the boundary, a relaxation mechanism is introduced along the external normal direction of the boundary via a Robin boundary condition on the chemical potential $\mu$: $\alpha \nabla_\bn \mu=\beta \mu_s-\mu$. This relaxation introduces an additional energy dissipation effect (term) to the energy dissipation rate:
\bena
\frac{d}{dt}E=-M_b^{(2)} \int_\Omega |\nabla \mu|^2 d\bx-M_s^{(2)} \int_{\partial \Omega} |\nabla_s \mu_s|^2 ds-\frac{M_b^{(2)}}{\alpha}\int_{\partial \Omega}(\beta \mu_s-\mu)^2ds. \label{1.18}
\eena
The time rate of change of the total mass in the bulk and on the surface follows the same equation as \eqref{mass-c0}.

Clearly, the dynamic boundary conditions in the above models are related, yielding different energy dissipation rates and mass conservation or transfer mechanisms between the bulk and the boundary. We briefly unwind the relations below.
\begin{itemize}
	\item  If $\beta  \to  \infty$ and $E_s=0$, \eqref{KLLM}-2 implies $\mu_s=0$ and $\nabla_\bn \phi=0$. Analogously, if $\alpha \to \infty$, \eqref{KLLM}-2 deuces $\nabla_\bn \mu=0$.  So, the  HNBC-CH Model can be viewed as a limiting case of the KLLM model.
	
	\item In  \eqref{KLLM}-2, $\beta \mu_s-\mu$ is explained as the difference of weighted surface chemical potential $\beta \mu_s$ and bulk chemical potential $\mu$. We define $\frac{1}{\alpha}$ as the relaxation rate, the system reaches the "equilibrium" at $\beta \mu_s=\mu$ when $\alpha \to 0$. So, the KLLM model reduces to the GMS model at $\alpha=0$.
	
	\item  If $\alpha \to \infty$, there is no relaxation and mass transfer between the boundary and bulk,  leading to  $\nabla_\bn \mu=0$. Thus, the KLLM model  reduces to the LW model.
\end{itemize}
In the analysis, we note that the KLLM model is a fairly general model which includes three others as special cases. However, it's not general enough to include the Gal Model and the  KL model. Recently, Hao Wu reviewed the derivation and analysis of the classical Cahn-Hilliard equation with static and dynamic boundary condition\cite{wu2021review}. In addition, all the above mentioned boundary conditions are valid in flat boundaries where the curvature vanishes. However, when the domain boundary has a non-negligible curvature, its can affect the dynamic boundary conditions. For arbitrarily shaped domain boundaries, we must include the curvature contribution to the boundary dynamics.  In many real-world applications, the boundary of the materials domain is not flat and it has non-negligible curvature. When the curvature effect is taken into account, the dynamic boundary conditions alluded to earlier must be modified to take into account the important geometric effect. This study will attempt to address this issue for a family of free energies.

In addition, the above reviewed phase field models are for purely dissipative systems. There are analogous phase field models for systems that allow both irreversible and reversible processes. One class of the phase field models includes the inertia effect to allow wave propagation \cite{galenko2005diffuse}. For example, the phase field model in \cite{cavaterra2014non} is given by
\ben
\phi_t=\nabla^2 \mu-\epsilon_0 \phi_{tt}, \quad \mu=-\nabla^2 \phi+f'(\phi)+\alpha_0 \phi_t, \quad \bx \in \Omega, \label{1.19}
\een
where $\epsilon_0$ is a measure of inertia and $\alpha_0>0$ is a viscosity coefficient.
Apparently, there is an underlying unified framework available to derive  models that are consistent with thermodynamical principles.

In this paper, we aim to develop such a general framework to  derive thermodynamically consistent models together with boundary conditions for nonequilibrium materials systems in any domains with piecewise smooth boundaries, where boundary dynamics and the boundary curvature effect are fully accounted for. We derive the general dynamic boundary condition as a constitutive relation by applying the generalized Onsager principle at the boundary analogous to what one does in the bulk.  We  elaborate on  two types of such dynamic boundary conditions specifically by prescribing two distinct energy dissipation mechanisms under the unified assumption that the mass flux at the boundary is dictated by the imbalance between the bulk chemical potential and the surface one. We show that most of the above reviewed boundary conditions are special cases of the two type boundary conditions. We illustrate the impact of boundary dynamics on the bulk structure using a phase field model for crystal growth in the end numerically.

The rest of the paper is organized as follows. In \S \ref{sec2}, we present a general model in domains with  smooth boundaries, whose free energy depends on gradients of the phase field variable up to the second order, and discuss its various limits. In \S\ref{sec3}, we discuss the extension to phase field models with a general free energy with high order spatial derivative and a nonlocal free energy.  In \S\ref{sec4}, we show the effect of surface dynamics on the crystal growth in a phase field model for crystal growth by numerical simulations. We summarize the results in \S\ref{sec5}.

\section{Thermodynamically consistent phase field models with  consistent dynamic boundary conditions} \label{sec2}

\noindent \indent We present a general framework for deriving transport equations and consistent dynamic boundary conditions for a phase field model that yields a negative   energy dissipation or a positive entropy production rate when  all dynamics are accounted for.  We illustrate the approach using the scalar phase field model for a binary material system with a free energy of  up to second order spatial derivatives of the phase field variable. Then, we elucidate the path for extending it to the more general free energy functional including the nonlocal free energy later. We make contact with the models mentioned in the introduction by examining limiting cases of the model and showing that many of those models are special cases of the general model. We discuss the derivation in the isothermal case in this paper so that the free energy is the proper potential to work with.

\subsection{Generalized Onsager principle} \label{sec2.1}

\noindent \indent The classical Onsager linear response theory on which the Onsager principle for dissipative systems is based provides a viable way to calculate dissipative forces in relaxation dynamics in an irreversible  nonequilibrium process \cite{onsager1931reciprocal1, onsager1931reciprocal2, onsager1953fluctuations}. In a general setting, the linear response theory states that given a chemical potential in an isothermal system, the generalized flux ${\phi_t}$ is proportional to the generalized force or chemical potential $\hat \mu$
\ben
{\phi_t}=-M \hat \mu, \label{2.1}
\een
where $M$ is called the mobility. In general, $M$ is an operator. For   dissipative systems where dynamics are irreversible, the additional Onsager reciprocal relation dictates that $M$ is symmetric; for conservative systems where dynamics are reversible, $M$ is antisymmetric \cite{Wang2020}. We note that when $M$ is a differential operator, like in the Cahn-Hilliard equation system, the property of $M$ is affected by the boundary conditions of the system. For a system where inertia is non-negligible and there coexist irreversible and reversible dynamics in the nonequilibrium process, we extend the force balance equation to a generalized Onsager principle \cite{yang2016hydrodynamic,Wang2020}
\ben
-M^{-1} {\phi_t}=\rho \phi_{tt}+\hat \mu \Leftrightarrow {\phi_t}=-M(\rho \phi_{tt}+\hat \mu), \label{GOP}
\een
where $\rho \phi_{tt}$ represents the inertia force,  $\rho$ is a measure of  mass and $M$ is the mobility operator which is not necessarily symmetric. We next use the generalized Onsager principle to derive the general phase field model along with its consistent boundary conditions for a binary material system.

\subsection{Models with free energy of up to second spatial derivatives} \label{sec2.2}

\noindent \indent Let the bulk free energy  in a fixed material domain $\Omega$ be given by
\ben
E_b[\phi]=\int_{\Omega} e_b(\phi, \nabla \phi, \nabla \nabla \phi) d\bx,\label{2.3}
\een
where $e_b$ is the energy density per unit volume. We consider a binary  material system with a boundary that may have its distinctive properties than the bulk and possesses its own surface energy of derivatives up to the second order in space
\ben
E_s[\phi]=\int_{\partial \Omega} e_s(\phi, \nabla_s \phi, \nabla_s \nabla_s \phi) ds, \label{2.4}
\een
where $e_s$ is the surface energy density per unit area, the phase field variable in the surface energy density is defined by
\ben\label{surface-phi}
\phi=\phi(\bx, t)|_{\partial \Omega}, \label{2.5}
\een
and $\nabla_s$ is the surface gradient operator over piecewise smooth boundary $\partial \Omega$ as that in the introduction. We note that
\eqref{surface-phi} is a critical assumption we adopt throughout the paper, which states that the phase field variable is continuous up to the boundary. The case where the phase field variable on the surface may not be the limit of the phase field variable in the bulk on the surface will be discussed in a sequel. Hence, we will not introduce a new notation for the  phase field variable   on the surface in this paper.

The free energy of the system is given by
\ben
E_{f}[\phi]=\int_{\Omega} e_b(\phi, \nabla \phi, \nabla \nabla \phi) d\bx+\int_{\partial \Omega} e_s(\phi, \nabla_s \phi, \nabla_s \nabla_s \phi) ds.\label{2.6}
\een

We add the kinetic energy in the bulk and on the boundary to account for the inertia effect in the system such that the total free energy is given by
\ben
E[\phi]=\int_{\Omega} [\frac{\rho}{2}{\phi_t}^2+e_b(\phi, \nabla \phi, \nabla \nabla \phi) d\bx+\int_{\partial \Omega} [\frac{\rho_s}{2} \phi_{t}^2+e_s(\phi, \nabla_s \phi, \nabla_s \nabla_s \phi)] ds,\label{2.7}
\een
where ${\phi_t}$ is the invariant  time derivative of $\phi$, $\rho$ and $\rho_s$ are two mass densities that measure the inertia in the bulk and on the surface, respectively.
We calculate the  time rate of change of the free energy as follows, assuming  domain $\Omega$ is fixed,
\ben
\bea{l}
\frac{dE}{dt}=\int_{\Omega} (\rho {\phi}_{tt}+\mu)  \phi_t d\bx+\int_{\partial \Omega} [\rho_s \phi_t \phi_{tt}+\frac{\partial e_s}{\partial \phi} \phi_t+\frac{\partial e_s}{\partial  \nabla_s \phi} \nabla_s \phi_t +\frac{\partial e_s}{\partial  \nabla_s \nabla_s \phi} \nabla_s \nabla_s \phi_t+\\
\qquad \bn \cdot \frac{\partial e_b}{\partial \nabla \phi} \phi_t+\frac{\partial e_b}{\partial \nabla \nabla \phi}: \bn \nabla \phi_t-\bn \nabla: \frac{\partial e_b}{\partial \nabla \nabla \phi} \phi_t ] ds\\
=\int_{\Omega} (\rho \phi_{tt}+\mu ) \phi_t d\bx+\int_{\partial \Omega} [\rho_s \phi_t \phi_{tt}+\frac{\partial e_s}{\partial \phi} \phi_t-\nabla_s \cdot \frac{\partial e_s}{\partial  \nabla_s \phi} \phi_t-2H \bn \cdot \frac{\partial e_s}{\partial  \nabla_s \phi} \phi_t \\
\qquad -\nabla_s \cdot \frac{\partial e_s}{\partial  \nabla_s \nabla_s \phi} \cdot \nabla_s \phi_t
-2H \bn \cdot \frac{\partial e_s}{\partial \nabla_s  \nabla_s \phi} \cdot \nabla_s \phi_t\\
\qquad +
\bn \cdot \frac{\partial e_b}{\partial \nabla \phi} \phi_t +\frac{\partial e_b}{\partial \nabla \nabla \phi}: \bn \nabla \phi_t-(\bn \nabla: \frac{\partial e_b}{\partial \nabla \nabla \phi}) \phi_t ] ds\\
=\int_{\Omega} (\rho\phi_{tt}+\mu)  \phi_t d\bx+\int_{\partial \Omega} [\rho_s \phi_t \phi_{tt}+\mu_s  \phi_t+\frac{\partial e_b}{\partial \nabla \nabla \phi}: \bn \bn (\bn \cdot \nabla \phi_t ) ] ds,
\eea\label{disp-eq}
\een
where $H$ is the mean curvature of the boundary, $\bn$ is the unit external normal of $\partial \Omega$, the bulk  and surface chemical potential are given respectively  by
\bena
\mu=\frac{\partial e_b}{\partial \phi}-\nabla \cdot \frac{\partial e_b}{\partial \nabla \phi}+\nabla \nabla : \frac{\partial e_b}{\partial \nabla \nabla \phi},\\
\mu_s=\frac{\partial e_s}{\partial \phi}-\nabla_s \cdot \frac{\partial e_s}{\partial  \nabla_s \phi}-2H \bn \cdot \frac{\partial e_s}{\partial  \nabla_s \phi}+\bn \cdot \frac{\partial e_b}{\partial \nabla \phi}-\bn \nabla : \frac{\partial e_b}{\nabla \nabla \phi}\\
\qquad+\nabla_s \nabla_s : \frac{\partial e_s}{\partial  \nabla_s \nabla_s \phi}+2H \bn \nabla_s: \frac{\partial e_s}{\partial \nabla_s  \nabla_s \phi} +
\nabla_s \cdot (2H\bn \cdot \frac{\partial e_s}{\partial \nabla_s  \nabla_s \phi})\\
\qquad+4H^2\bn \bn :\frac{\partial e_s}{\partial \nabla_s  \nabla_s \phi}-\nabla_s\bn:\frac{\partial e_b}{\partial \nabla \nabla \phi} -2H\bn \bn :\frac{\partial e_b}{\partial \nabla \nabla \phi}. \label{2.9}
\eena
Note that (i) the surface chemical potential includes contributions from the surface free energy as well as that from the bulk energy confined to the boundary;  (ii)
the mean curvature shows up in the surface chemical potential, indicating that curvature of the boundary affects the dynamics at the boundary; and (iii) more surface terms can appear if the free energy density function depends on higher order spatial derivatives. We adopt the Einstein notation for tensors, denote tensor product of vector $\bn$ and $\bv$ as $\bn\bv=n_i v_j$,  use one dot $\cdot$ to represent inner product $\bn\cdot \bv=n_i v_i$ and two dots $:$ to represent contraction of two second order tensor $\bA:\bC=A_{ij} C_{ij}$, where $\bn,\bv$ are vectors, $\bA, \bC$ are second order tensors, and the Einstein notation is adopted.

\subsubsection{Dynamics in the bulk} \label{sec2.2.1}

\noindent \indent We apply the generalized Onsager principle firstly to the bulk integral in \eqref{disp-eq} to obtain the transport equation for $\phi$ in $\Omega$
\ben
-M_b^{-1}\phi_t=\rho \phi_{tt}+\mu  \Leftrightarrow \phi_t=-M_b(\rho \phi_{tt}+\mu), \quad \bx \in \Omega, \label{bulk-eq}
\een
where $M_b$ is the mobility operator and $M_b^{-1}$ is the friction operator which is positive semi-definite to ensure energy dissipation.  We consider the mobility operator in the following form in this study
\ben
M_b=M_b^{(1)}-\nabla \cdot \bM_b^{(2)} \cdot \nabla, \label{2.11}
\een
where $M_b^{(1)}\geq 0$ is a scalar function of $\phi$ and $ \bM_b^{(2)} \in \bR^{3\times 3}$ is a semi-definite positive matrix which can be a function of $\phi$ as well. If $ \bM_b^{(2)}=M_b^{(2)}\bI$ and $M_b^{(2)}$ is also a scalar function of $\phi$, then such a special case $\nabla \cdot \bM_b^{(2)} \cdot \nabla=\nabla \cdot (M_b^{(2)} \nabla)$ can be obtained. We note that the derivation applies to a more general mobility operator with high order derivatives as well, which we will not pursue in this study. The presence of spatial derivatives in the mobility indicates the nonlocal interaction is accounted for in the friction operator $M_b^{-1}$. This is shown in the form of pseudo-differential operators.
With this, the energy dissipation rate   reduces to
\ben
\bea{l}
\frac{dE}{dt}=- \int_{\Omega} [(\mu+\rho \phi_{tt}) M_b^{(1)} (\mu+\rho \phi_{tt})+\nabla (\mu+\rho \phi_{tt})  \cdot \bM_b^{(2)} \cdot \nabla (\mu+\rho \phi_{tt})] d\bx\\
+\int_{\partial \Omega} [(\mu_s+\rho_s \phi_{tt}) \phi_t+\mu_c \nabla_\bn \phi_t+(\mu +\rho \phi_{tt})\bn \cdot \bM_b^{(2)} \cdot \nabla (\mu+\rho \phi_{tt}) ] ds, \label{energy-diss}
\eea
\een
where $\mu_c=\frac{\partial e_b}{\partial \nabla \nabla \phi}: \bn \bn$. We denote the generalized chemical potential in the bulk by $\tilde \mu=\mu+\rho \phi_{tt}$ and in the surface by $\tilde \mu_s=\mu_s+\rho_s \phi_{tt},$ respectively.  We remark that $\bn \cdot \bM_b^{(2)} \cdot \nabla (\mu+\rho \phi_{tt})$ is the inward mass flux across the boundary. This  physical quantity is determined by the balance between the surface and bulk chemical potential. We next derive thermodynamically consistent boundary conditions.

\subsubsection{Dynamics on boundaries} \label{sec2.2.2}

\noindent \indent We first recognize that the boundary energy flux density is a quadratic form and then apply the Onsager principle the second time to the energy flux density to establish a dynamic constitutive equation at the boundary:
\ben
\left (
\bea{l}
\phi_t\\
f_m\\
\nabla_{\bn} \phi_t
\eea
\right)
=-M_{3\times 3} \cdot
\left (
\bea{l}
\tilde \mu_s\\
\tilde \mu\\
\mu_c
\eea
\right), \label{2.13}
\een
where $f_m=\bn \cdot  \bM_b^{(2)} \cdot \nabla\tilde \mu $ is the inward mass flux and $M_{3\times 3}\geq 0$ is the surface mobility operator, a $3\times 3$ matrix or second order tensor. $M_{3\times3}\geq0$ means that its symmetric operator is semi-positive definite. Then,
\ben
\bea{l}
\frac{dE}{dt}=- \int_{\Omega} [\tilde \mu M_b^{(1)} \tilde \mu+\nabla \tilde \mu  \cdot \bM_b^{(2)} \cdot \nabla \tilde \mu] d\bx\\
\qquad-
\int_{\partial \Omega} [( \tilde \mu_s, \tilde \mu, \mu_c) (M_{3\times 3} ) ( \tilde \mu_s, \tilde \mu, \mu_c)^T] ds \leq 0, \label{energy-diss12}
\eea
\een
which indicates the system is dissipative.
We examine two special cases that include most of the models and boundary conditions mentioned in the introduction below.

\noindent {\bf Case 1: a purely dissipative boundary condition.}

We define a symmetric mobility operator as follows
\ben
M_{3\times 3}=
\left (
\bea{lcr}
M_s+\frac{\beta^2}{\alpha} & -\frac{\beta}{\alpha} & 0\\
-\frac{\beta}{\alpha} & \frac{1}{\alpha} & 0
\\
0 & 0 & M_c
\eea
\right),\label{purely}
\een
where $M_s$ is a semi-definite positive operator, $\alpha\geq 0 $ is a friction coefficient,
$1/\beta\geq 0$ is a length parameter, and $M_c$ is a semi-definite positive mobility operator. We note that these mobility operators can be differential operators. When $M_c$ is a scalar in a simple example and $M_c \to \infty$, $\mu_c=\frac{\partial e_b}{\partial \nabla \nabla \phi}: \bn \bn=0$; while $M_c \to 0$, $\nabla_\bn \phi_t=0$.

This constitutive equation establishes a balance between the inward mass flux  at the boundary and the generalized chemical potential difference between the bulk and the surface: it assumes the inward mass flux is proportional to the  difference between the chemical potential in the bulk and the weighted one at the boundary. When the weighted surface energy is higher than the bulk energy confined to the boundary, the mass flux is  inward; otherwise, the mass flux flows outward. In either case, the total energy dissipates when $M_{3\times 3}\geq 0$.

The governing equation together with the boundary conditions in this model is given as follows
\bena\label{g-model1}
\parl{\phi}{t}=-M_b\tilde \mu,  \quad \bx \in \Omega, \\
\parl{\phi}{t}=-(M_s+\frac{\beta^2}{\alpha})\tilde \mu_s+\frac{\beta }{\alpha}\tilde \mu,  \quad \bx \in \partial \Omega,\\
\alpha \bn \cdot \bM_b^{(2)} \cdot \nabla \tilde \mu=-\tilde \mu+\beta \tilde \mu_s, \quad
\nabla_\bn \phi_t=-M_c \mu_c,  \quad \bx \in \partial \Omega,\\
\phi(0)=\phi_0, \quad \bx \in \Omega\cup\partial \Omega. \label{gover1}
\eena

The corresponding energy dissipation rate  is given by
\ben
\bea{l}
\frac{dE}{dt}
=- \int_{\Omega} [\tilde \mu M_b^{(1)} \tilde \mu+\nabla \tilde \mu  \cdot \bM_b^{(2)} \cdot \nabla \tilde \mu ]d\bx-\int_{\partial \Omega} [\mu_c M_c \mu_c ]ds\\
\qquad +\int_{\partial \Omega} [\tilde \mu_s (\phi_t+ \beta  \bn \cdot \bM_b^{(2)} \cdot  \nabla  \tilde \mu )] ds
-\frac{1}{\alpha}\int_{\partial \Omega} (\beta \tilde \mu_s-\tilde \mu)^2ds
.
\eea\label{energy-diss1}
\een
The surface transport equation of $\phi$ can be rewritten into an alternative form on boundary $\partial \Omega$
\ben
\parl{\phi}{t}=-M_s\tilde \mu_s-\beta f_m, \quad \bx \in \partial \Omega, \label{2.18}
\een
where
\ben
f_m=\bn\cdot \bM_b^{(2)} \cdot \nabla \tilde {\mu}=\frac{1}{\alpha}(\beta \tilde \mu_s-\tilde \mu). \label{2.19}
\een
This indicates that the time rate of change of mass fraction $\phi$ is proportional to the outward mass flux and the generalized surface chemical potential.

\noindent{\bf Case 2: a dissipative and transportive boundary condition}

In the second case, we propose another mobility operator as follows
\ben
M_{3\times 3}=
\left (
\bea{lcr}
M_s & \frac{\beta}{\alpha} & 0\\
-\frac{\beta}{\alpha} & \frac{1}{\alpha} & 0
\\
0 & 0 & M_c
\eea
\right)=
\left (
\bea{lcr}
M_s & 0 & 0\\
0 & \frac{1}{\alpha} & 0\\
0 & 0 & M_c
\eea
\right)+
\left (
\bea{lcr}
0 & \frac{\beta}{\alpha} & 0\\
-\frac{\beta}{\alpha} & 0 & 0\\
0 & 0 & 0
\eea
\right),\label{nonpurely}
\een
which includes an antisymmetric component, contributing to transport dynamics at the boundary, in addition to the  positive semi-definite operator in $M_{3\times 3}$. The antisymmetric mobility component represents an energy exchange between the bulk and the boundary without inducing any dissipation.

The governing equation together with the boundary conditions in this model is summarized as follows
\bena\label{g-model2}
\parl{\phi}{t}=-M_b\tilde \mu,  \quad \bx \in \Omega, \\
\parl{\phi}{t}=-M_s \tilde \mu_s-\frac{\beta}{\alpha} \tilde \mu, \quad \bx \in \partial \Omega,\\
\alpha \bn \cdot \bM_b^{(2)} \cdot \nabla \tilde \mu=-\tilde \mu+\beta \tilde \mu_s, \quad
\nabla_\bn \phi_t=-M_c \mu_c,  \quad \bx \in \partial \Omega,\\
\phi(\bx, 0)=\phi_0, \quad \bx \in \Omega\cup\partial \Omega. \label{gover2}
\eena
Notice that the mobility matrix has an antisymmetric component which does not contribute to the energy dissipation. This set of boundary conditions  has the following  interpretation: the time rate of change of the phase field variable at the boundary is proportional to both the surface chemical potential and the inward flux across the boundary.

The corresponding  energy dissipation rate is given by
\ben
\bea{l}
\frac{dE}{dt}=- \int_{\Omega} [\tilde \mu M_b^{(1)} \tilde \mu+\nabla \tilde \mu  \cdot \bM_b^{(2)} \cdot \nabla \tilde \mu] d\bx-
\int_{\partial \Omega} [\tilde \mu_s M_s \tilde \mu_s+
\frac{1}{\alpha}\tilde \mu^2+\mu_c M_c \mu_c ]ds.\label{energy-diss2}
\eea
\een

The two types of dynamic boundary conditions are derived from two different considerations of the mobility operator, which contribute to distinct energy dissipation mechanisms, following the generalized Onsager principle under a unified assumption that the boundary mass flux is proportional to the difference of the bulk energy confined at the boundary and a weighted surface energy. In the first case, the time rate of change of mass fraction ( phase field variable) is proportional to the surface chemical potential and the outward mass flux. As a result, the  distinctive surface energy dissipation rate is directly linked to the magnitude of the mass flux across the boundary surface. In the second case, the time rate of change of  mass fraction  is proportional to the surface chemical potential and the inward mass flux. Consequentially, the distinctive surface energy dissipation rate  is measured by the bulk   chemical potential confined to the surface. Two different dissipative mechanisms define two  different dynamic models at the boundary. There are more cases that one can elaborate on by specifying specific form of operator $M_{3\times 3}$, which we will not enumerate  in this study.

In general, mobility operator in the bulk  $M_b=M_b^{sym}+M_b^{anti}$  in \eqref{GOP} can be decomposed into symmetric and antisymmetric part, where $M_b^{sym}$ is semi-definite positive. There can be many more thermodynamically consistent boundary conditions that are compatible to the given bulk transport equation.
For the mobility operators at the boundary, we consider the following forms, analogously to the bulk,
\ben
M_s=M_s^{(1)}-\nabla_s \cdot \bM_s^{(2)}\cdot \nabla_s,\quad
M_c=M_c^{(1)}-\nabla_s \cdot \bM_c^{(2)} \cdot \nabla_s, \label{Mc}
\een
where $M_c^{(1)}\geq 0$, $M_s^{(1)}\geq 0$, and $\bM_s^{(2)}$ and $\bM_c^{(2)}$ are $3\times 3$ positive semi-definite matrices. Then,
\bena
-\int_{\partial \Omega} [\tilde \mu_s M_s \tilde \mu_s] ds=-\int_{\partial \Omega}[\tilde \mu_s M_s^{(1)}\tilde \mu_s+
\nabla_s \tilde \mu_s \cdot \bM_s^{(2)} \cdot \nabla_s \tilde \mu_s]ds \\
-\int_{\partial \Omega} [2H\tilde \mu_s \bn\cdot \bM_s^{(2)} \cdot \nabla_s \tilde \mu_s] ds, \label{ms}
\eena
\bena
-\int_{\partial \Omega} [\mu_c M_c \mu_c] ds=-\int_{\partial \Omega}[\mu_c M_c^{(1)}\mu_c+\nabla_s \mu_c \cdot \bM_c^{(2)} \cdot \nabla_s \mu_c]ds\\
-\int_{\partial \Omega} [2H\mu_c \bn\cdot \bM_c^{(2)} \cdot \nabla_s \mu_c] ds.\label{mc}
\eena
Whether or not the energy dissipation rate at the boundary is nonpositive depends on the last terms in \eqref{ms} and \eqref{mc}, which are linearly proportional to the mean curvature.

If $\bM_s^{(2)}={M}_s^{(2)} \bI$ and $\bM_c^{(2)}={M}_c^{(2)} \bI$, where ${M}_s^{(2)}\geq 0$ and $M_c^{(2)}\geq 0$ are scalar functions of $\phi$, the last terms in \eqref{ms} and \eqref{mc} vanish and the energy dissipation rate at the boundary are nonpositive due to $\bn \cdot \nabla_s \mu_s=\bn \cdot \nabla_s \mu_c=0$. Of course, $H=0$ is also a sufficient condition for non-positive energy dissipation rates.

It follows from  \eqref{gover1} that
\bena
\frac{d}{dt}[\int_{\Omega} \beta \phi d\bx+\int_{\partial \Omega } \phi ds]\\
=-\beta \int_{\Omega}M_b^{(1)} \tilde \mu d\bx-\int_{\partial \Omega} [ M_s^{(1)} \tilde \mu_s+2H\bn \cdot \bM_s^{(2)}\cdot \nabla_s \tilde \mu_s] ds. \label{2.25}
\eena
If $\bM_s^{(2)}={M}_s^{(2)} \bI$ or $H=0$,
\ben
\frac{d}{dt}[\int_{\Omega} \beta \phi d\bx+\int_{\partial \Omega } \phi ds]
=-\beta \int_{\Omega}M_b^{(1)} \tilde \mu d\bx-\int_{\partial \Omega}  M_s^{(1)} \tilde \mu_s ds.\label{2.26}
\een
If $M_b^{(1)}=0, M_s^{(1)}=0$,
\ben
\frac{d}{dt}[\int_{\Omega} \beta \phi d\bx+\int_{\partial \Omega } \phi ds]=0. \label{2.27}
\een
This indicates a weighted mass in the bulk and the mass over the surface is conserved under this dynamic boundary condition.
In this case, parameter  $\beta$ in the dynamic boundary condition can be interpreted as the weighted mass at the bulk to that over the surface.

Model  \eqref{g-model1} and \eqref{g-model2} give a general  phase field model with two different dynamic boundary conditions, where the surface transport equation of $\phi$ at the boundary sets the two models apart.   In the first one, the across boundary mass flux contributes directly to the energy dissipation on the surface; while in the second, it is the bulk chemical potential limited to the boundary contributes to the energy dissipation on the surface directly.  We next examine various limiting cases to show that most models mentioned in the introduction are special cases of model \eqref{g-model1} and \eqref{g-model2}, respectively. Specifically, when the across boundary mass flux is forbidden, i.e., $\alpha=\infty$ and $\beta=0$, we show that the two types of dynamic boundary conditions are  identical.

\subsection{Limiting cases } \label{sec2.3}

\noindent \indent
Since \eqref{g-model1} and \eqref{g-model2} describe two different   dynamics at the boundary, we  examine them  closely in several limiting cases and make contact with the models alluded to in the introduction and in the literature.
Notice that when $\rho=\rho_s=0$, the phase field model reduces to the over-damped limit where the inertia force is neglected,  $\tilde \mu=\mu$ and $\tilde \mu_s=\mu_s$. We present limits of the under-damped case in the following, the results for the over-damped ones are obtained by setting $\rho=\rho_s=0$.

\begin{itemize}
	\item   Let $\beta  \to  0$, the boundary conditions reduce to
	\ben
	\alpha \bn \cdot \bM_b^{(2)} \cdot \nabla \tilde \mu=-\tilde \mu, \quad \phi_t =-M_s \tilde \mu_s, \quad \nabla_\bn \phi_t=-M_c \mu_c, \quad \bx \in \partial \Omega. \label{2.28}
	\een
	The first equation states that the mass flux between the bulk and the boundary is completely dictated by the bulk chemical potential extrapolated (or confined) to the boundary. The second one shows the relaxation dynamics of mass fraction at the surface are dictated completely by the surface chemical potential. This indicates that the across boundary mass flux does not interfere with the surface dynamics at the boundary. The third one is necessary only when the free energy has second order spatial derivatives, which represents the relaxation dynamics of the directional derivative of the volume fraction.
	
	Then the energy dissipation rate reduces to
	\bena
	\frac{dE}{dt}=- \int_{\Omega} [\tilde \mu M_b^{(1)} \tilde \mu+\nabla \tilde \mu \cdot \bM_b^{(2)} \cdot \nabla \tilde \mu] d\bx
	-\int_{\partial \Omega} [\tilde \mu_s M_s \tilde \mu_s] ds\\
	- \int_{\partial \Omega} [\frac{1}{\alpha}\tilde \mu^2  +\mu_c M_c \mu_c] ds \leq 0.\label{2.29}
	\eena
	Model \eqref{g-model1} and \eqref{g-model2} are identical  and dissipative in this limit.
	
	
	\item Limit $\alpha \to 0$  is a singular limit. We will conduct the limiting process in the following order. Firstly, we substitute \eqref{gover1}-3 into \eqref{gover1}-2; secondly, we take the limit $\alpha\to 0$ in \eqref{gover1}-3. The results are given by
	\bena
	\phi_t+\beta \bn \cdot \bM_b^{(2)} \cdot \nabla \tilde \mu =-M_s \tilde \mu_s,\quad \bx \in  \partial  \Omega,\\
	\beta \tilde \mu_s=\tilde \mu, \quad \nabla_\bn \phi_t=-M_c \mu_c, \quad \bx \in  \partial  \Omega.\label{2.30}
	\eena
	These conditions state that the bulk chemical potential and the weighted surface one reach a balance at the boundary and the time rate of change in the phase field variable at the boundary  is given by the rate of change due to the outward mass flux and the surface chemical potential.
	The energy dissipation rate reduces to
	\bena
	\frac{dE}{dt}=- \int_{\Omega} [\tilde \mu M_b^{(1)} \tilde \mu +\nabla \tilde \mu   \cdot \bM_b^{(2)} \cdot \nabla \tilde \mu  ]d\bx
	-\int_{\partial \Omega} [\tilde \mu_s  M_s \tilde \mu_s  ] ds\\
	\qquad
	-\int_{\partial \Omega} [\mu_c M_c \mu_c ]ds.\label{disp6}
	\eena
	The model is dissipative with the boundary conditions.
	
	If we take the singular limit in \eqref{gover2}, we end up with
	\ben
	\tilde \mu=0, \quad  \tilde \mu_s=0, \quad \nabla_\bn \phi_t=-M_c \mu_c, \quad \bx \in  \partial  \Omega. \label{2.32}
	\een
	and the energy dissipation rate is given by
	\bena
	\frac{dE}{dt}=- \int_{\Omega} [\tilde \mu  M_b^{(1)} \tilde \mu +\nabla \tilde \mu   \cdot \bM_b^{(2)} \cdot \nabla \tilde \mu  ]d\bx
	-\int_{\partial \Omega} [\mu_c M_c \mu_c ]ds.\label{disp7}
	\eena
	The two models do not give the same set of boundary conditions. They indeed describe two different dynamics in the limit.
	
	\item When $\alpha \to \infty$, we have
	\ben
	\phi_t=-M_s \tilde \mu_s,\quad  \bn \cdot \bM_b^{(2)} \cdot \nabla \tilde \mu=0, \quad \nabla_\bn \phi_t=-M_c \mu_c, \quad \bx \in  \partial  \Omega, \label{2.34}
	\een
	and the energy dissipation rate is given by \eqref{disp6}. The boundary conditions stipulate that the mass flux at the boundary vanishes and the phase field dynamics at the boundary is dictated by relaxation dynamics of the surface mass fraction exclusively.
	The two models are once again identical and dissipative.

\end{itemize}

We summarize the dynamic boundary conditions and energy dissipation rates in table \ref{table1} and \ref{table2} in the limits. In the case $\alpha \rightarrow 0$, the two models are not identical, whereas they are the same in the other two cases.  
While both $\alpha$ and $\beta$ are finite, we consider  the limiting cases with respect to boundary mobility operators $M_s$ and $M_c$, respectively. The results are tabulated in tables
\ref{table3} and \ref{table4}, respectively. Regardless what are the surface mobility operators, the two boundary conditions are different in the cases, yielding two distinct thermodynamically consistent phase field models with consistent dynamic boundary conditions. Note that the two dynamic boundary conditions differ in how the mass flux transfers across the boundary and how mass fraction dynamics on the surface are prescribed. In table \ref{table5}, we demonstrate that most  existing models mentioned in the introduction are special cases of the general model with the first type dynamic boundary condition. It is obvious that none of these models are the special case of the general model with the second type dynamic boundary condition. In fact, there are also some progresses in the nonlocal model with dynamic  boundary conditions recently \cite{knopf2021nonlocal}, which is also a special case of the general model with first type of dynamic boundary condition. We discuss the general nonlocal models with two types of dynamic boundary conditions in the next section.
Notice that  models  \eqref{g-model1} and \eqref{gover2} include all the models alluded to in the introduction as special cases except for the KL model  when the free energy is limited to functionals of up to the first order spatial derivative. This is shown clearly in Table \ref{table5}. Thus, the phase field
model presented here is indeed a general phase field model. Moreover, the boundary conditions derived in this study include the curvature effect for an arbitrarily shaped piecewise smooth boundary, which have  not been considered in previous studies. Finally, we note that any set of boundary conditions delineated here or their limiting cases may appear as boundary conditions on a smooth  piece of the piecewise smooth boundary so that the overall boundary conditions can be a combination of the sets of dynamic boundary conditions.
\begin{table}[!htbp] 
	\vskip 12 pt
	\centering
	\scriptsize
	{
		\begin{tabular}{|c|c|c|}
			\hline
			Case &   I &   II \\
			\hline
			
			$\beta \to 0$ & \multicolumn{2}{|c|} {$\parl{\phi}{t}=-M_s\tilde \mu_s$,\quad
				$\alpha \bn \cdot \bM_b^{(2)} \cdot \nabla \tilde \mu=-\tilde  \mu$,\quad
				$\nabla_\bn \phi_t=-M_c\mu_c$} \\
			\hline
			$\alpha \to 0$ & {$\parl{\phi}{t}+\beta \bn \cdot M_b^{(2)}\cdot \tilde \mu=-M_s\tilde  \mu_s$, \quad
				$\beta \tilde  \mu_s=\tilde \mu$, \quad
				$\nabla_\bn \phi_t=-M_c\mu_c$}&{$\tilde \mu=0, \quad  \tilde \mu_s=0, \quad \nabla_\bn \phi_t=-M_c \mu_c$}\\
			\hline
			$\alpha \to \infty$ & \multicolumn{2}{|c|} {$\parl{\phi}{t}=-M_s\tilde  \mu_s$, \quad
				$\bn \cdot \bM_b^{(2)} \cdot \nabla \tilde  \mu=0,$ \quad
				$\nabla_\bn \phi_t=-M_c\mu_c$}\\
			\hline
	\end{tabular}}
	\caption{Dynamical boundary conditions in three limiting cases. }  \label{table1}
\end{table}

\begin{table}[!htbp] 
	\vskip 10 pt
	\centering
	\scriptsize
	{
		\begin{tabular}{|c|c|}
			\hline
			Case &   I  \\
			\hline
			$\beta \to 0$&  {$\frac{dE}{dt}=- \int_{\Omega} [\tilde \mu M_b^{(1)} \tilde \mu+\nabla \tilde \mu  \cdot \bM_b^{(2)} \cdot \nabla \tilde \mu ]d\bx-\int_{\partial \Omega} [\tilde \mu_s M_s \tilde \mu_s ] ds-\frac{1}{\alpha}\int_{\partial \Omega} [ {\tilde \mu}^2]ds
				-\int_{\partial \Omega} [\mu_c M_c \mu_c ]ds$}\\
			\hline
			$\alpha \to 0$ &{$\frac{dE}{dt}=- \int_{\Omega} [\tilde \mu M_b^{(1)} \tilde  \mu+\nabla \tilde  \mu  \cdot \bM_b^{(2)} \cdot \nabla \tilde  \mu ]d\bx
				-\int_{\partial \Omega} [\tilde \mu_s M_s \tilde \mu_s +\mu_c M_c \mu_c ]ds$}\\
			\hline
			$\alpha \to \infty$ &{$\frac{dE}{dt}=- \int_{\Omega} [\tilde \mu M_b^{(1)} \tilde \mu+\nabla \tilde \mu  \cdot \bM_b^{(2)} \cdot \nabla \tilde \mu ]d\bx-\int_{\partial \Omega} [\tilde \mu_s M_s \tilde \mu_s ] ds
				-\int_{\partial \Omega} [\mu_c M_c \mu_c ]ds$}\\
			\hline
				Case &  II \\
			\hline
			$\beta \to 0$& {$\frac{dE}{dt}=- \int_{\Omega} [\tilde \mu M_b^{(1)} \tilde \mu+\nabla \tilde \mu  \cdot \bM_b^{(2)} \cdot \nabla \tilde \mu ]d\bx-\int_{\partial \Omega} [\tilde \mu_s M_s \tilde \mu_s ] ds-\frac{1}{\alpha}\int_{\partial \Omega} [ {\tilde \mu}^2]ds
				-\int_{\partial \Omega} [\mu_c M_c \mu_c ]ds$}\\
			\hline
			$\alpha \to 0$ &{$\frac{dE}{dt}=- \int_{\Omega} [\tilde \mu  M_b^{(1)} \tilde \mu +\nabla \tilde \mu   \cdot \bM_b^{(2)} \cdot \nabla \tilde \mu  ]d\bx
								-\int_{\partial \Omega} [\mu_c M_c \mu_c ]ds$}\\
			\hline
			$\alpha \to \infty$ &{$\frac{dE}{dt}=- \int_{\Omega} [\tilde \mu M_b^{(1)} \tilde \mu+\nabla \tilde \mu  \cdot \bM_b^{(2)} \cdot \nabla \tilde \mu ]d\bx-\int_{\partial \Omega} [\tilde \mu_s M_s \tilde \mu_s ] ds
				-\int_{\partial \Omega} [\mu_c M_c \mu_c ]ds$}\\
			\hline
	\end{tabular}} 
	\caption{Energy dissipation rates in three  limiting  cases. 
		$\rm I$ and $\rm II$ represent the first and second type of boundary condition, respectively, when they are different.} \label{table2}
\end{table}

\begin{table}[!htbp]
	\vskip 12 pt
	\centering
	\scriptsize
	{
		\begin{tabular}{|c|c|}
			\hline
			Case &  I \\
			\hline
			$M_s \to 0$ &{$\parl{\phi}{t}=-\frac{\beta^2}{\alpha} \tilde \mu_s+\frac{\beta }{\alpha}\tilde \mu$, \quad
				$\alpha \bn \cdot \bM_b^{(2)} \cdot \nabla \tilde  \mu=\beta \tilde \mu_s-\tilde \mu$, \quad
				$\nabla_\bn \phi_t=-M_c\mu_c$} \\
			\hline
			$M_s \to \infty$ &{$\parl{\phi}{t}=-\beta  \bn \cdot \bM_b^{(2)} \cdot \nabla \tilde  \mu$, \quad
				$\tilde \mu_s=0$, \quad
				$\nabla_\bn \phi_t=-M_c\mu_c$} \\
			\hline
			$M_c \to 0$ &{$\parl{\phi}{t}=-(M_s+\frac{\beta^2}{\alpha}) \tilde \mu_s+\frac{\beta }{\alpha}\tilde  \mu$, \quad
				$\alpha \bn \cdot \bM_b^{(2)} \cdot \nabla \tilde \mu=\beta \tilde \mu_s-\tilde \mu$, \quad
				$\nabla_\bn \phi_t=0$} \\
			\hline
			$M_c \to \infty$ &{$\parl{\phi}{t}=-(M_s+\frac{\beta^2}{\alpha}) \tilde \mu_s+\frac{\beta }{\alpha} \tilde \mu$, \quad
				$\alpha \bn \cdot \bM_b^{(2)} \cdot \nabla \tilde \mu=\beta \tilde \mu_s-\tilde \mu$, \quad
				$\mu_c=0$} \\
			\hline
				Case &    II \\
			\hline
			$M_s \to 0$ &{$\parl{\phi}{t}=-\frac{\beta}{\alpha} \tilde \mu$, \quad $\alpha \bn \cdot \bM_b^{(2)} \cdot \nabla \tilde  \mu=\beta \tilde \mu_s-\tilde \mu$, \quad $\nabla_\bn \phi_t=-M_c\mu_c$} \\
			\hline
			$M_s \to \infty$ &{$\parl{\phi}{t}=\beta \bn \cdot \bM_b^{(2)} \cdot \nabla \tilde  \mu$, \quad
				$\tilde  \mu_s=0$,  \quad $\nabla_\bn \phi_t=-M_c\mu_c$} \\
			\hline
			$M_c \to 0$ &{$\parl{\phi}{t}=-M_s \tilde \mu_s-\frac{\beta}{\alpha} \tilde \mu$, \quad
				$\alpha \bn \cdot \bM_b^{(2)} \cdot \nabla \tilde  \mu=\beta  \tilde \mu_s-\tilde \mu$, \quad
				$\nabla_\bn \phi_t=0$} \\
			\hline
			$M_c \to \infty$ &{$\parl{\phi}{t}=-M_s \tilde \mu_s-\frac{\beta}{\alpha} \tilde  \mu$, \quad
				$\alpha \bn \cdot \bM_b^{(2)} \cdot \nabla \tilde \mu=\beta \tilde  \mu_s-\tilde  \mu$, \quad
				$\mu_c=0$} \\
			\hline
	\end{tabular}}  
	\caption{Two types of dynamic boundary conditions. }\label{table3}
\end{table}
\begin{table}[!htbp]
	\vskip 12 pt
	\centering
			\scriptsize
	{
		\begin{tabular}{|c|c|}
			\hline
			Case &  I \\
			\hline
			$M_s \to 0$ &{$\frac{dE}{dt}=- \int_{\Omega} [\tilde \mu M_b^{(1)}\tilde  \mu+\nabla \tilde \mu  \cdot \bM_b^{(2)} \cdot \nabla \tilde \mu ]d\bx-\frac{1}{\alpha}\int_{\partial \Omega} [ (\beta \tilde \mu_s-\tilde \mu)^2]ds
				-\int_{\partial \Omega} [\mu_c M_c \mu_c ]ds$}\\
			\hline
			$M_s \to \infty$ &{$\frac{dE}{dt}=- \int_{\Omega} [\tilde \mu M_b^{(1)} \tilde \mu+\nabla \tilde \mu  \cdot \bM_b^{(2)} \cdot \nabla \tilde \mu ]d\bx -\frac{1}{\alpha}\int_{\partial \Omega} [ \tilde \mu^2]ds
				-\int_{\partial \Omega} [\mu_c M_c \mu_c ]ds$}\\
			\hline
			$M_c \to 0$ &{$\frac{dE}{dt}=- \int_{\Omega} [\tilde \mu M_b^{(1)} \tilde \mu+\nabla \tilde \mu  \cdot \bM_b^{(2)} \cdot \nabla \tilde \mu ]d\bx-\int_{\partial \Omega} [\tilde \mu_s M_s \tilde \mu_s ] ds-\frac{1}{\alpha}\int_{\partial \Omega} [ (\beta \tilde \mu_s-\tilde \mu)^2]ds$}\\
			\hline
			$M_c \to \infty$ &{$\frac{dE}{dt}=- \int_{\Omega} [\tilde \mu M_b^{(1)} \tilde \mu+\nabla \tilde \mu  \cdot \bM_b^{(2)} \cdot \nabla \tilde \mu ]d\bx-\int_{\partial \Omega} [\tilde \mu_s M_s \tilde \mu_s ] ds-\frac{1}{\alpha}\int_{\partial \Omega} [ (\beta \tilde \mu_s-\tilde \mu)^2]ds$}\\
			\hline
			Case &  II \\
			\hline
			$M_s \to 0$ &{$\frac{dE}{dt}=- \int_{\Omega} [\tilde \mu M_b^{(1)} \tilde \mu+\nabla \tilde \mu  \cdot \bM_b^{(2)} \cdot \nabla \tilde \mu ]d\bx-\frac{1}{\alpha}\int_{\partial \Omega} [ \tilde \mu^2]ds
				-\int_{\partial \Omega} [\mu_c M_c \mu_c ]ds$}\\
			\hline
			$M_s \to \infty$ &{$\frac{dE}{dt}=- \int_{\Omega} [\tilde \mu M_b^{(1)} \tilde \mu+\nabla \tilde \mu  \cdot \bM_b^{(2)} \cdot \nabla \tilde \mu ]d\bx-\frac{1}{\alpha}\int_{\partial \Omega} [ \tilde \mu^2]ds
				-\int_{\partial \Omega} [\mu_c M_c \mu_c ]ds$}\\
			\hline
			$M_c \to 0$ &{$\frac{dE}{dt}=- \int_{\Omega} [\tilde \mu M_b^{(1)} \tilde \mu+\nabla \tilde \mu  \cdot \bM_b^{(2)} \cdot \nabla \tilde \mu ]d\bx-\int_{\partial \Omega} \tilde \mu_s M_s \tilde \mu_s ds-\frac{1}{\alpha}\int_{\partial \Omega} [ \tilde \mu^2]ds$}\\
			\hline
			$M_c \to \infty$ &{$\frac{dE}{dt}=- \int_{\Omega} [\tilde \mu M_b^{(1)} \tilde \mu+\nabla \tilde \mu  \cdot \bM_b^{(2)} \cdot \nabla \tilde \mu ]d\bx-\int_{\partial \Omega}\tilde  \mu_s M_s \tilde \mu_s ds-\frac{1}{\alpha}\int_{\partial \Omega} [ \tilde \mu^2]ds$}\\
			\hline
	\end{tabular}}
	\caption{The  free energy dissipation rates corresponding to the two types of boundary dynamics. }\label{table4}
\end{table}
\begin{table}[!htbp]
	\vskip 12 pt
	\centering
			\scriptsize
	{
		\begin{tabular}{|c|c|}
			\hline
			Models &  Specific conditions for the model with first type of dynamic boundary condition  \\
			\hline
			Gal model\cite{gal2006cahn} &{$M_b=M_b^{(2)}\nabla^2$, \quad
				$M_s=1$, \quad
				$\alpha=0$, \quad $\rho=\rho_s=\mu_c=M_c=H=0$} \\
			\hline
			GMS model\cite{goldstein2011cahn} &{$M_b=M_b^{(2)}\nabla^2$, \quad
				$M_s=M_s^{(2)}\nabla_s^2$, \quad
				$\alpha=0$, \quad $\rho=\rho_s=\mu_c=M_c=H=0$} \\
			\hline
			LW model\cite{liu2019energetic} &{$M_b=M_b^{(2)}\nabla^2$, \quad
				$M_s=M_s^{(2)}\nabla_s^2$, \quad
				$\alpha\to \infty$, \quad $\rho=\rho_s=\mu_c=M_c=H=0$} \\
			\hline
			KLLM model\cite{knopf2020phase} &{$M_b=M_b^{(2)}\nabla^2$, \quad
				$M_s=M_s^{(2)}\nabla_s^2$, \quad $\rho=\rho_s=\mu_c=M_c=H=0$} \\
			\hline
	\end{tabular}}
		\caption{Relation between the  general model with first types of dynamic boundary conditions and several existing models in the literature. } \label{table5}
\end{table}
\clearpage

\subsection{Mixed dynamic boundary conditions} \label{sec2.4}

\noindent\indent Let us consider a domain $\Omega$ with piecewise smooth boundaries $\partial \Omega =\cup_{i=1}^N{\Gamma_i}$, where $\Gamma_i$ and $\Gamma_j$ are either mutually disjoint or adjacent smooth surfaces, $i,j=1,\cdots, N$. The energy dissipation rate in \eqref{energy-diss} can be rewritten into
\ben
\bea{l}
\frac{dE}{dt}=- \int_{\Omega} [\tilde \mu M_b^{(1)} \tilde \mu+\nabla \tilde \mu  \cdot \bM_b^{(2)} \cdot \nabla \tilde \mu] d\bx\\
\qquad   +\sum_i^N \int_{\Gamma_i} [\tilde \mu_s \phi_t+\mu_c \bn\cdot \nabla \phi_t+\tilde \mu \bn \cdot \bM_b^{(2)} \cdot \nabla \tilde \mu ] ds.\label{disp-piece}
\eea
\een
For the surface terms in \eqref{disp-piece}, either of two boundary conditions \eqref{gover1} and \eqref{gover2} can be implemented. We illustrate this for case $N=2$.

The following boundary conditions give dissipative boundary conditions to the above energy dissipation rate:
\bena\label{g-model21}
\parl{\phi}{t}=-(M_s+\frac{\beta_1^2}{\alpha_1}) \tilde \mu_s+\frac{\beta_1 }{\alpha_1}\tilde \mu,  \quad\bx \in \Gamma_1,\\ 
\alpha_1 \bn \cdot \bM_b^{(2)} \cdot \nabla \tilde \mu=-\tilde \mu+\beta_1 \tilde \mu_s,  \quad\bx \in \Gamma_1, \\ \nabla_\bn \phi_t=-M_c \mu_c,  \quad\bx \in \Gamma_1,\\
\parl{\phi}{t}=-M_s \tilde \mu_s-\frac{\beta_2}{\alpha_2} \tilde \mu, \quad \bx \in \Gamma_2,\\
\alpha_2 \bn \cdot \bM_b^{(2)} \cdot \nabla \tilde \mu=-\tilde \mu+\beta_2 \tilde \mu_s,\quad \bx \in \Gamma_2,\\
 \nabla_\bn \phi_t=-M_c \mu_c,
\quad \bx \in \Gamma_2. \label{2.36}
\eena
From \eqref{energy-diss}, we have
\ben
\bea{l}
\frac{dE}{dt}
=- \int_{\Omega} [\tilde \mu M_b^{(1)} \tilde \mu+\nabla\tilde \mu  \cdot \bM_b^{(2)} \cdot \nabla \tilde \mu ]d\bx
+\int_{\Gamma_1} [\tilde \mu_s (\phi_t+ \beta_1  \bn \cdot \bM_b^{(2)} \cdot  \nabla  \tilde \mu )] ds\\
\qquad
-\frac{1}{\alpha_1}\int_{\Gamma_1} (\beta_1 \tilde \mu_s-\tilde \mu)^2ds
-\int_{\Gamma_1} [\mu_c M_c \mu_c ]ds
+\int_{\Gamma_2} [\tilde \mu_s (\phi_t+\frac{\beta_2 }{\alpha_2} \tilde \mu )] ds\\
\qquad
-\frac{1}{\alpha_2}\int_{\Gamma_2} [\tilde \mu^2]ds
-\int_{\Gamma_2} [\mu_c M_c\mu_c ]ds.
\eea\label{energy-diss-mixed}
\een
It   is straightforward to generalize it to cases where $N>2$. We note that the function space for the solution of the initial boundary value problems  should be chosen such that weak derivatives exist in the bulk and on the surface. We will not elaborate on this in this paper.

\subsection{Examples} \label{sec2.5}

\noindent \indent We present a few free energy functionals for binary phase field models describing immiscible, miscible polymeric materials, molecular beam epitaxy (MBE)  and crystal growth, respectively.

\subsubsection{Polynomial double-well and Flory-Huggins mixing free energy for multiphase polymers} \label{sec2.5.1}

\noindent \indent
We consider a general free energy functional involving polynomial double well or Flory-Huggins bulk mixing energy and the conformational entropy in both the bulk and on the surface
\bena
E_b=\int_{\Omega}[ \frac{\rho}{2}\phi_t^2+\frac{\gamma_1}{2}\nabla \phi \cdot {\bf D}\cdot \nabla \phi +\gamma_2 f(\phi)] d\bx, \\
 E_s=\int_{\partial \Omega}[ \frac{\rho_s}{2}\phi_t^2+\frac{\zeta_1}{2}\nabla_s \phi \cdot {\bf D}_s \cdot \nabla_s \phi+\zeta_2 g(\phi)] ds, \label{double}
\eena
where $\phi$ is a phase variable vector, ${\bf D}$ and ${\bf D}_s$ are the positive semi-definite anisotropic coefficients of the conformational entropy in the bulk and surface \cite{warren1995prediction,karma1999phase,zhao2017numerical}, respectively, $\gamma_i, \zeta_i, i=1,2$ are parameters.
The corresponding  dynamic governing equations  are given  by setting $\mu_c=M_c=0$ in \eqref{gover1} and \eqref{gover2}.

\subsubsection{Free energy for molecular beam epitaxy} \label{sec2.5.2}

\noindent \indent
A bulk and a surface free energy for molecular beam epitaxy (MBE) are similar to \eqref{double}, but with different choices of energy densities  $f$ and $g$ \cite{li2003thin,yang2017numerical}:
\bena
E_b=\int_{\Omega}[ \frac{\rho}{2}\phi_t^2+\frac{\gamma_1}{2}\nabla \phi \cdot {\bf D}\cdot \nabla \phi +\gamma_2 f(\phi)] d\bx, \\
 E_s=\int_{\partial \Omega}[ \frac{\rho_s}{2}\phi_t^2+\frac{\zeta_1}{2}\nabla_s \phi \cdot {\bf D}_s \cdot \nabla_s \phi+\zeta_2 g(\phi)] ds. \label{2.39}
\eena
There are two choices of $(\phi)f$ and $g(\phi)$ in MBE models, one is
\ben
f(\phi)=\frac{1}{4}(1-|\nabla \phi|^2)^2, \bx \in \Omega, \quad g(\phi)=\frac{1}{4}(1-|\nabla \phi|^2)^2, \bx \in \partial \Omega, \label{2.40}
\een
with slope selection and  the other  is
\ben
f(\phi)=-\frac{1}{2}\ln (1+|\nabla \phi|^2), \bx \in \Omega, \quad  g(\phi)=-\frac{1}{2}\ln (1+|\nabla \phi|^2), \bx \in \partial \Omega, \label{2.41}
\een
without slope selection.  Setting $\mu_c=M_c=M_b^{(2)}=0$, we obtain the desired dynamic equations from \eqref{gover1} and \eqref{gover2}.

\subsubsection{Free energy for crystal growth models} \label{sec2.5.3}

\noindent \indent A bulk free energy for the phase field crystal growth model is given by
\bena
E_b=\int_{\Omega} [\frac{\rho}{2}\phi_t^2+\frac{\phi}{2}(-\varepsilon+(\nabla ^2+1)^2)\phi+\frac{\phi^4}{4}]\mathrm{d {\bx}}, \label{2.42}
\eena
where $\phi$ represents an atomistic density field, which is the deviation of the density from the average density and is a conserved field variable, $\epsilon$ is a parameter related to the temperature, that is, higher $\epsilon$ corresponds to a lower temperature \cite{elder2004modeling,elder2007phase}.

Likewise, we propose the following for the surface energy
\ben
E_s=\int_{\partial \Omega}[\frac{\rho_s}{2}\phi_t^2+\frac{1}{2}(\nabla_s ^2\phi)^2- |\nabla_s\phi|^2+g(\phi)] ds, \label{2.43}
\een
where $g(\phi)$ is a prescribed surface energy density.
The corresponding governing equations, boundary conditions and energy dissipation rates are given by \eqref{gover1} and \eqref{gover2} with the free energy functionals substituted, respectively.
We could also consider an anisotropic phase-field crystal model by using anisotropic conformational entropy in the bulk and the surface energy functional \cite{kundin2017application}.

\section{Nonlocal models and other extensions} \label{sec3}

\noindent \indent We now consider  phase field models with a nonlocal free energy \cite{giacomin1997phase,gal2018doubly}, where the free energy is given by
\ben
E_b=\int_{\Omega}[\int_{\Omega} \frac{1}{4}J(\|\bx-\by\|) (\phi(\bx, t)-\phi(\by, t))^2 d\by+  f(\phi)] d\bx, \label{3.1}
\een
where $J(\|\bx\|)$ is the interaction kernel and $f$ is the free energy density for the bulk. We assume the interaction between the bulk and the boundary has been built in the interaction kernel. This form of the free energy is perhaps more generic than the one that depends on spatial derivatives of the phase variable.

The chemical potential is calculated as follows
\ben
\mu
=\int_{\Omega} J(\|\bx-\by\|) (-\phi(\by,t)) d\by+f'(\phi)+a(\bx)\phi(\bx,t),\label{3.2}
\een
where $a(\bx)=\int_{\Omega} J(\|\bx-\by\|) d\by.$
Likewise, we consider  the surface energy given by
\ben
E_s=\int_{\partial \Omega}[\int_{\partial \Omega} \frac{1}{4}K(\|\bx-\by\|) (\phi(\bx, t)-\phi(\by, t))^2 ds_{\by} +  g(\phi)] ds_{\bx},\label{3.3}
\een
where $g$ is the surface energy density.
The surface chemical potential is calculated as follows
\bena
\mu_s=\int_{\partial \Omega} K(\|\bx-\by\|) (\phi(\bx,t)-\phi(\by,t)) ds_{\by}+g'(\phi)\\
\quad=\int_{\partial \Omega} K(\|\bx-\by\|) (-\phi(\by,t)) ds_{\by}+g'(\phi)+a_S(\bx)\phi(\bx,t),\label{3.4}
\eena
where $a_S(\bx)=\int_{\partial \Omega} K(\|\bx-\by\|) ds_{\by}.$
The total free energy, including the inertia effect, is then given by
\ben
E=E_b+E_s+\int_{\Omega} \frac{\rho}{2}( \phi_t)^2d\bx +\int_{\partial \Omega} \frac{\rho_s}{2}(\phi_t)^2 ds. \label{3.5}
\een
We calculate the time rate of change of the free energy
\ben
\frac{d}{dt}E=\int_{\Omega} \tilde \mu \phi_t d\bx+\int_{\partial \Omega} \tilde \mu_s \phi_t ds_{\bx}. \label{3.6}
\een
We apply the Onsager principle to the bulk term to arrive at
\ben
\phi_t=-M_b\tilde \mu,\quad \bx \in \Omega, \label{3.7}
\een
where $M_b$ is the mobility operator. For $M_b=M_b^{(1)}-\nabla \cdot \bM_b^{(2)} \cdot \nabla$,
\ben
\bea{l}
\frac{d}{dt}E=-\int_{\Omega} [\tilde \mu M_b^{(1)} \tilde \mu+\nabla \tilde \mu  \cdot \bM_b^{(2)} \cdot \nabla \tilde \mu] d\bx+\int_{\partial \Omega} [\tilde \mu_s \phi_t +\tilde \mu \bn \cdot \bM_b^{(2)} \cdot \nabla \tilde \mu ]dS_{\bx}\\
=-\int_{\Omega} [\tilde \mu M_b^{(1)} \tilde \mu+\nabla \tilde \mu  \cdot \bM_b^{(2)} \cdot \nabla \tilde \mu] d\bx+\int_{\partial \Omega} [\tilde \mu_s \phi_t +\tilde \mu f_m ]ds.\label{3.8}
\eea
\een
We propose boundary condition as follows
\ben
\left(
\bea{l}
\phi_t\\
f_m
\eea\right)
=-M_{2\times 2} \cdot \left (
\bea{l}
\tilde \mu_s \\
\tilde \mu\eea
\right),
\quad \bx \in \partial \Omega, \label{3.9}
\een
where $M_{2\times 2}\geq 0$ is the boundary mobility operator.

The energy dissipation rate is given by
\ben
\bea{l}
\frac{d}{dt}E=-\int_{\Omega} [\tilde \mu M_b^{(1)} \tilde \mu+\nabla \tilde \mu \cdot \bM_b^{(2)}\cdot \nabla \tilde \mu] d\bx-\int_{\partial \Omega} [(\tilde \mu_s, \tilde \mu)\cdot M_{2\times 2} \cdot (\tilde \mu_s, \tilde \mu)^T]ds. \label{3.10}
\eea
\een
By specifying $M_{2\times 2}$ as the $2\times 2$ upper left  sub-matrix in $M_{3\times 3}$ in the previous section, we arrive at two types of dynamic boundary conditions analogous to the above mentioned, which we will not repeat them here. We note that the authors derived a nonlocal model with the first type of dynamic boundary conditions and proved the weak and strong well-posedness of the system recently in \cite{knopf2021nonlocal} although they used two distinct phase field variables for the bulk and surface respectively.

The dynamic boundary conditions for the nonlocal model are similar to the ones with weakly nonlinear interactions through high order derivatives except that the chemical potential and the time rate of change of the phase field variable at the boundary do not have the explicit dependence on the bulk chemical potential.  The explicit connection between the bulk and the surface is in fact established through the nonlocal kernel in the free energy in the bulk effectively.


In the previous section, we present the results for a free energy with up to the second order spatial derivatives. This  can be readily extended to include more general free energy functionals. It requires one to consider physically i) what would be the appropriate boundary conditions when the variation of the free energy is carried; ii) when the mobility operator includes high order spatial differential operators, how to deal with the boundary terms  generated while applying integrations by parts in the context of thermodynamical consistency. The method presented here should be able to guide the generalization to those cases straightforwardly.


\section{Numerical results}\label{sec4}

\noindent \indent In this section, we use the crystal growth model alluded to earlier, as an example, to illustrate the effect of dynamic boundary conditions to the solution in the bulk numerically.  We adopt the energy quadratization (EQ) and finite difference method to discretize the governing equation of the phase field crystal growth model \cite{jing2019second,jing2020linear}. We assume the
dimensionless bulk free energy and surface energy in a fixed rectangle domain are given respectively  by
\bena
E_b=\int_{\Omega} [\frac{|\nabla^2 \phi|^2}{2}-|\nabla \phi|^2+\frac{1-\epsilon}{2}\phi^2+\frac{\phi^4}{4}]\mathrm{d {\bx}},\\
E_s=\int_{\partial \Omega} [\frac{|\nabla^2_s \phi|^2}{2}-|\nabla_s \phi|^2+\frac{1-\epsilon_s}{2}\phi^2+\frac{\phi^4}{4}]\mathrm{d {s}}, \label{4.1}
\eena
where $\epsilon, \epsilon_s$ are positive constant parameters.
The corresponding chemical potentials in the bulk and on the boundary are calculated as follows
\ben
\bea{l}
\mu=\nabla^4 \phi+2\nabla^2 \phi+(1-\varepsilon)\phi+\phi^3,\quad \mu_c=\nabla \nabla \phi :\bn \bn,\\
\mu_s=\nabla_s^4 \phi+2\nabla_s^2 \phi+(1-\varepsilon_s)\phi+\phi^3-\nabla^3 \phi \cdot \bn-2\nabla \phi \cdot \bn-\nabla_s \bn: \nabla \nabla \phi. \label{4.2}
\eea
\een

\noindent \indent  We present some numerical examples of the crystal growth model with dynamic boundary conditions on a part of the boundary. We use a 2D computational domain, in which the four sides are labeled as $\Gamma_1, \Gamma_2, \Gamma_3$ and $\Gamma_4$, respectively. Dynamics in the bulk is governed by
\bena
\parl{\phi}{t}=M_b^{(2)}\nabla^2 \mu, \label{4.3}
\eena
where $M_b^{(2)}$ is a positive constant, while dynamic boundary conditions on each boundary are given respectively by
\bena
\parl{\phi}{t}=-(-M_s^{(2)}\nabla_s^2+\frac{\beta_1^2}{\alpha_1}) \mu_s+\frac{\beta_1 }{\alpha_1}\mu, \quad
\alpha_1 M_b^{(2)}\bn \cdot   \nabla \mu=-\mu+\beta_1 \mu_s, \quad\bx \in \Gamma_1\\
 \nabla_\bn \phi_t=M_c ^{(2)}\nabla^2\mu_c,  \quad\bx \in \Gamma_1,\quad
\bn \cdot \nabla \phi=\bn \cdot \nabla^3 \phi=\bn \cdot \nabla \mu=0, \quad \bx \in \Gamma_2, \Gamma_3,  \Gamma_4.\label{4.4}
\eena
Namely, we allow dynamic boundary conditions on one side of the boundary and static boundary conditions on the rest.

We use the energy quadratization technique together with the Crank-Nicolson method in time, and the second order finite difference method on staggered grids in space to derive a thermodynamically consistent numerical algorithm by introducing two intermediate scalar variables in the bulk and on the surface, respectively.   The numerical algorithm guarantees that the total energy  dissipates in time and space \cite{jing2022DBC}.
Simulations of the crystal growth model with static, homogeneous Neumann boundary conditions uniformly along the boundary, corresponding to the case of a zero surface energy,  can be found in \cite{jing2020linear}.

In all simulations, solid crystallites with Hexagonal ordering in 2D is initially placed in the centre of the  computational domain, which is assigned an average density $\overline{\phi}$. The initial condition is given by
\bena
\phi_0(\bf r)=\overline{\phi}+w({\bf r})(A\phi_s(\bf r)),\label{4.5}
\eena
where
\bena
w(\bf r)=\begin{cases} (1-(\frac{|\bf r-\bf r_0| }{\bf d_0})^2)^2& \text{if }  \frac{|\bf r-\bf r_0|}{\bf d_0}\le 1, \\
	0& \text{otherwise}, \end{cases} \label{4.6}
\eena
\bena
\phi_s ({\bf r})=\cos(\frac{q}{\sqrt{3}}y)\cos(q x)-\frac{1}{2}\cos(\frac{2q}{\sqrt 3}y), \label{4.7}
\eena
${\bf r}=(x,y)$, $\bf r_0$ is the center coordinate of  the domain, and $\bf d_0$ is $\frac{1}{6}$ of the domain length in the x-direction. The domain is given by  $ \Omega =[0,\frac{2\pi}{q}a] \times [0,\frac {\sqrt {3} \pi}{q}b]$, $a=10$ and $b=12$. The other parameter values are  $\varepsilon=0.325, \overline{\phi}=\frac{\sqrt {\varepsilon}}{2}, A=\frac{4}{5}(\overline{\phi}+\frac{\sqrt{15\varepsilon-36\overline {\phi}^2}}{3})$ and $q=\frac{\sqrt{3}}{2}.$ For the initial  condition on the surface and the surface energy at the boundary, we set $\epsilon_s=\epsilon$ and $\phi=A\phi_s$ on $\Gamma_1$ as the first initial  condition at the surface. We move the initial condition to the right by $6h_x$ as the second initial condition of $\phi$ on the boundary to check the potential grain boundary effects induced by the surface energy, where $h_x$ is the spatial step size. For simplicity, we call these two initial  conditions on the surface as ordered and shifted  initial condition on the surface, respectively. Besides these, we set $M_b^{(2)}=M_s^{(2)}=1$ and $M_c^{(2)}=0$ in the following simulations.

\begin{figure*}
	\centering
	\subfigure[]{
		\begin{minipage}[b]{0.3\linewidth}
			\includegraphics[width=1\linewidth]{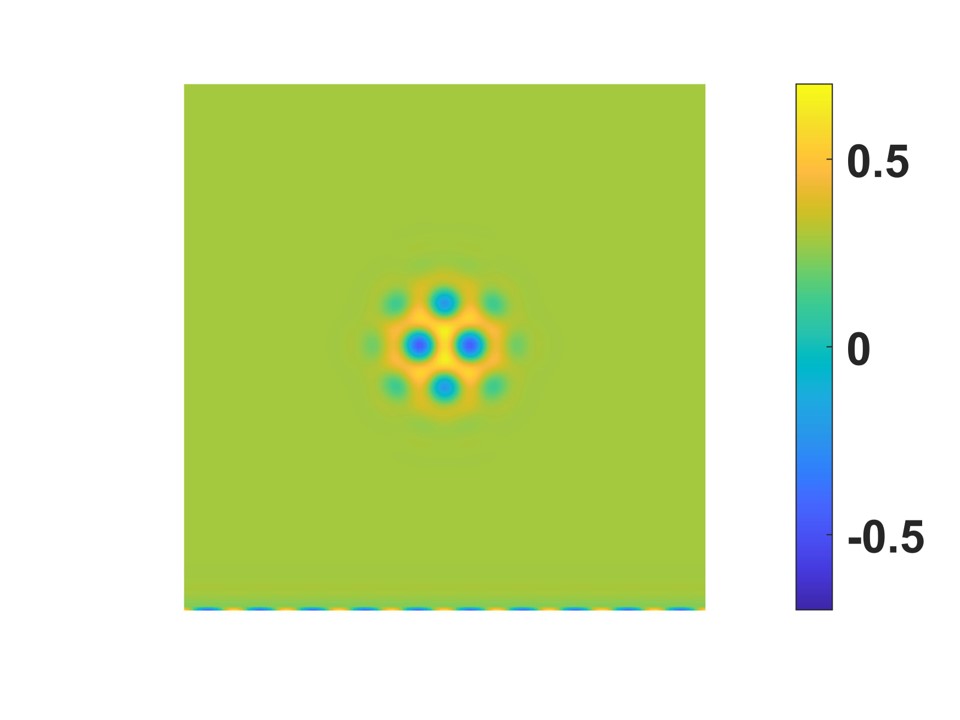}
	\end{minipage}}
	\subfigure[]{
		\begin{minipage}[b]{0.3\linewidth}
			\includegraphics[width=1\linewidth]{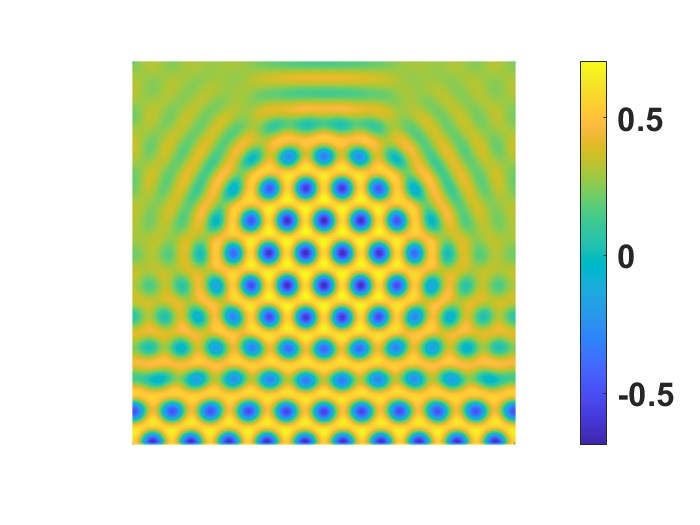}
	\end{minipage}}
	\subfigure[]{
		\begin{minipage}[b]{0.3\linewidth}
			\includegraphics[width=1\linewidth]{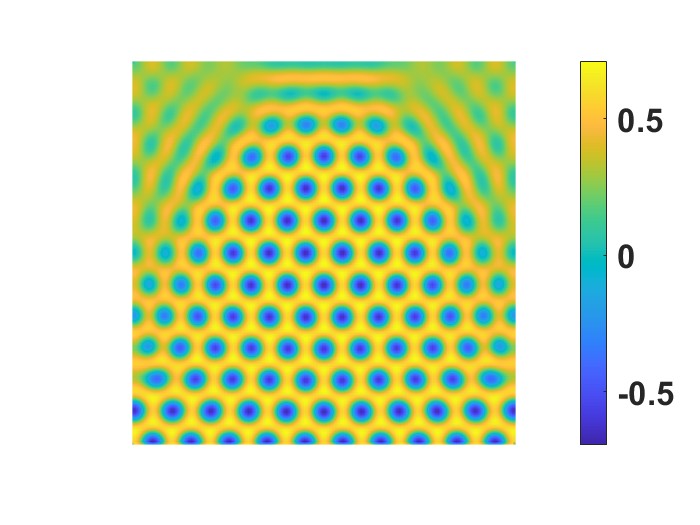}
	\end{minipage}}
	\subfigure[]{
		\begin{minipage}[b]{0.3\linewidth}
			\includegraphics[width=1\linewidth]{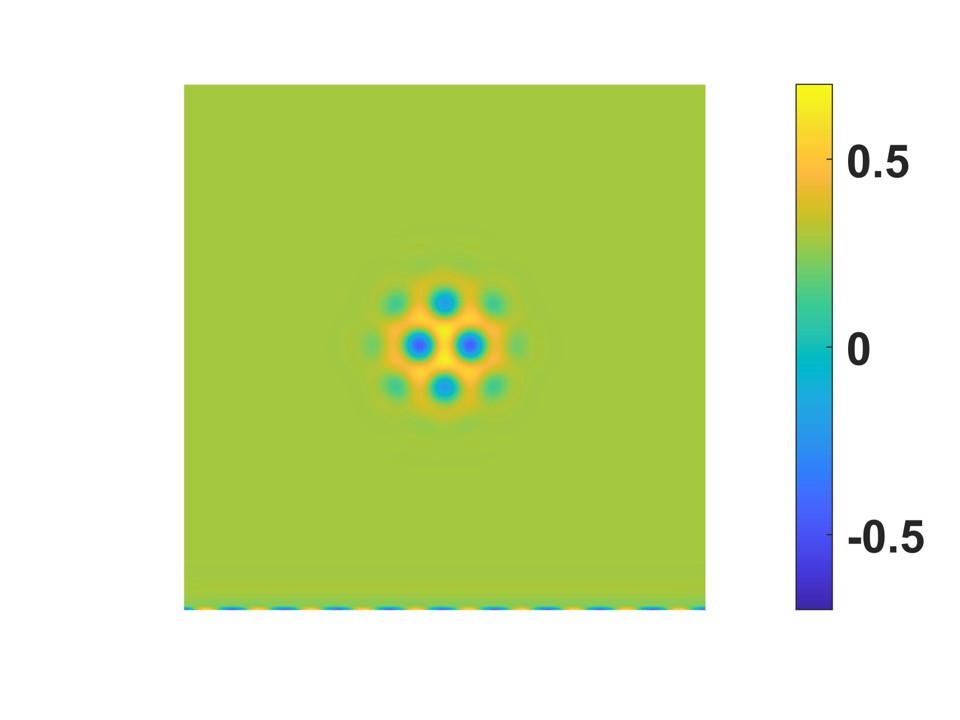}
	\end{minipage}}
	\subfigure[]{
		\begin{minipage}[b]{0.3\linewidth}
			\includegraphics[width=1\linewidth]{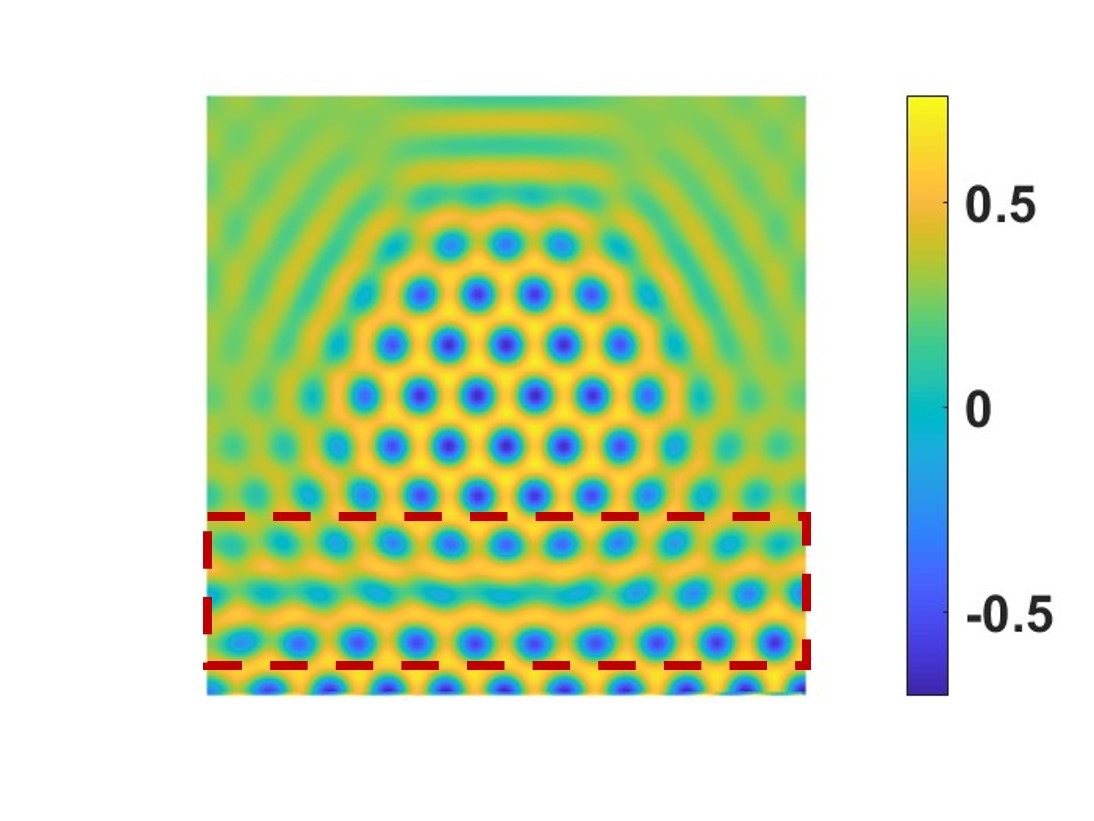}
	\end{minipage}}
	\subfigure[]{
		\begin{minipage}[b]{0.3\linewidth}
			\includegraphics[width=1\linewidth]{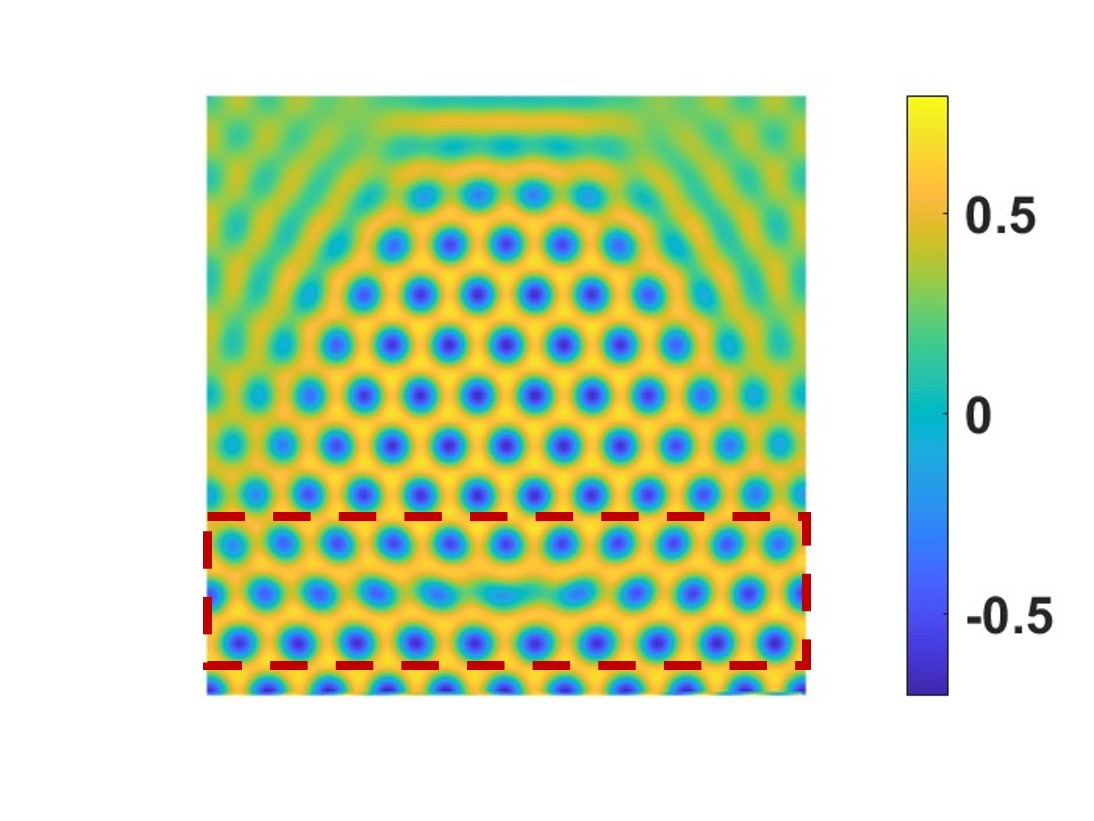}
	\end{minipage}}
	\caption{ The ordered growth vs the grain boundary effect induced by a shifted initial boundary  condition in (d)-(f).   Snapshots of the numerical solution of $\phi$ are taken  with $128 \times 128$ meshes in 2D space at $T=0, 32, 40$, respectively. Time step  $\delta t=1 \times 10^{-2}$ and $\alpha_1=\beta_1=1$ are used  in the simulations. (a)-(c): $\phi=A\phi_s$ is used at boundary $\Gamma_1$; (d)-(f): the   shifted initial boundary condition is used in the simulation, where the grain boundary effect was shown in the highlighted region. }\label{figure1}
\end{figure*}

In the following, we investigate the effect of the surface energy on the bulk structure by varying two parameters  $\alpha_1, \beta_1$.  At first, the ordered and shifted initial conditions on the surface are depicted in Figure \ref{figure1}-(a,d), respectively. Figure \ref{figure1}-(a-c) show that the crystal grows from the bulk and the surface simultaneously with the ordered initial condition without a grain boundary effect. Figure \ref{figure1}-(d-f) show the grain boundary effect induced by the shifted initial condition on the surface in the highlighted region. This simulation demonstrates that a dynamic boundary condition can significantly affect crystal growth in the bulk.

\begin{figure*}
	\centering
	\subfigure[]{
		\begin{minipage}[b]{0.3\linewidth}
			\includegraphics[width=1\linewidth]{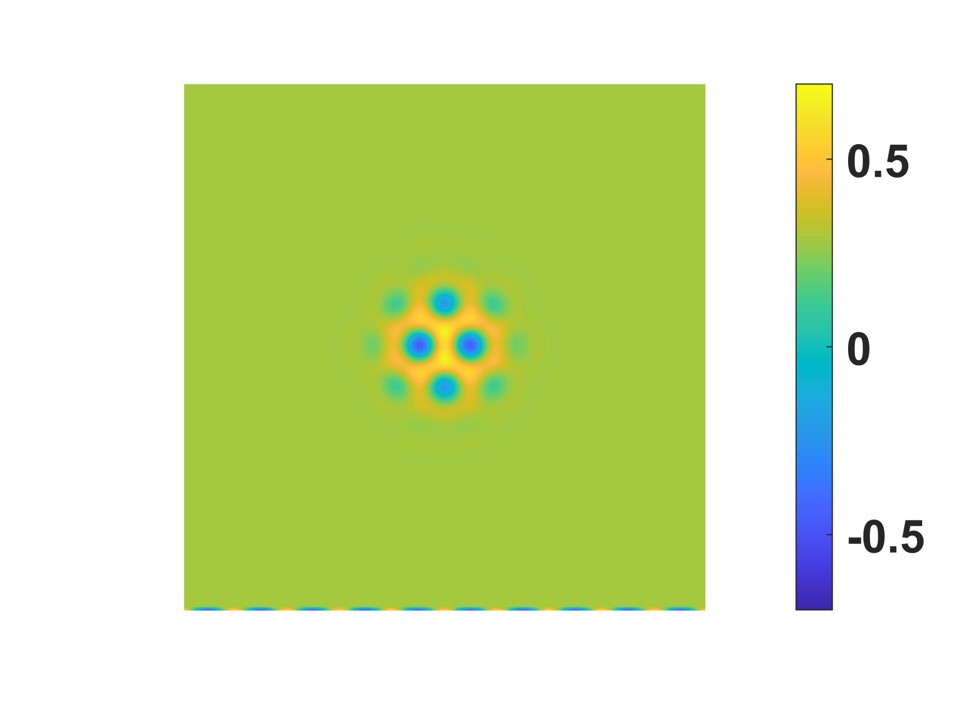}
	\end{minipage}}
	\subfigure[]{
		\begin{minipage}[b]{0.3\linewidth}
			\includegraphics[width=1\linewidth]{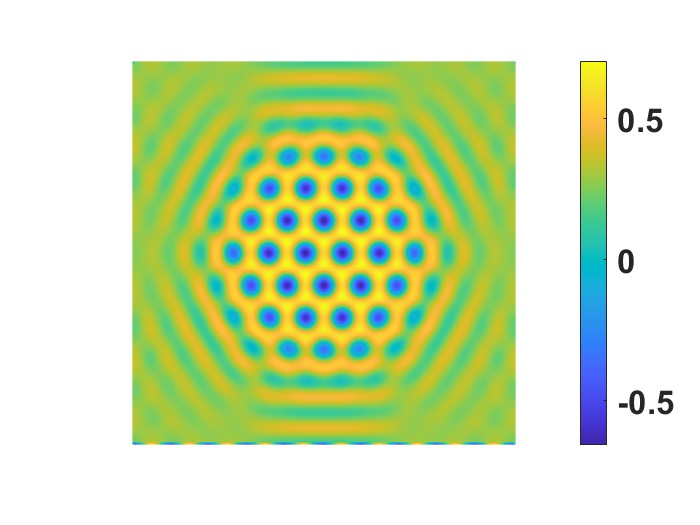}
	\end{minipage}}
	\subfigure[]{
		\begin{minipage}[b]{0.3\linewidth}
			\includegraphics[width=1\linewidth]{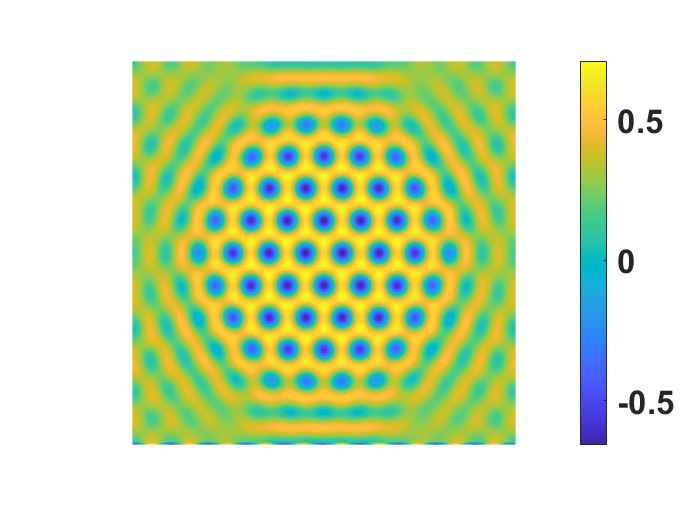}
	\end{minipage}}
	\subfigure[]{
		\begin{minipage}[b]{0.3\linewidth}
			\includegraphics[width=1\linewidth]{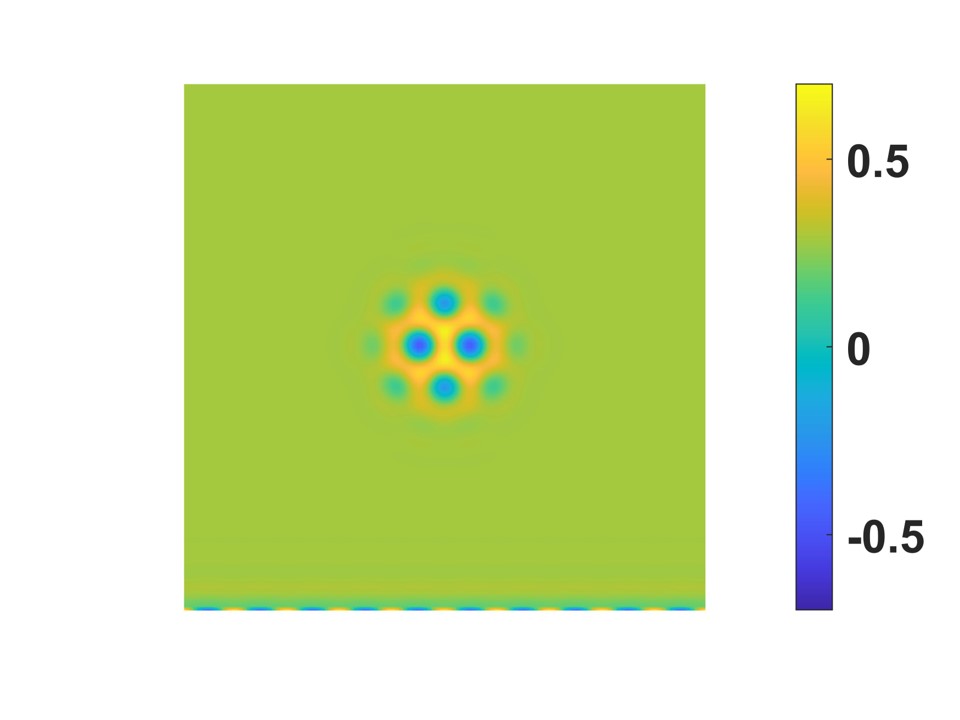}
	\end{minipage}}
	\subfigure[]{
		\begin{minipage}[b]{0.3\linewidth}
			\includegraphics[width=1\linewidth]{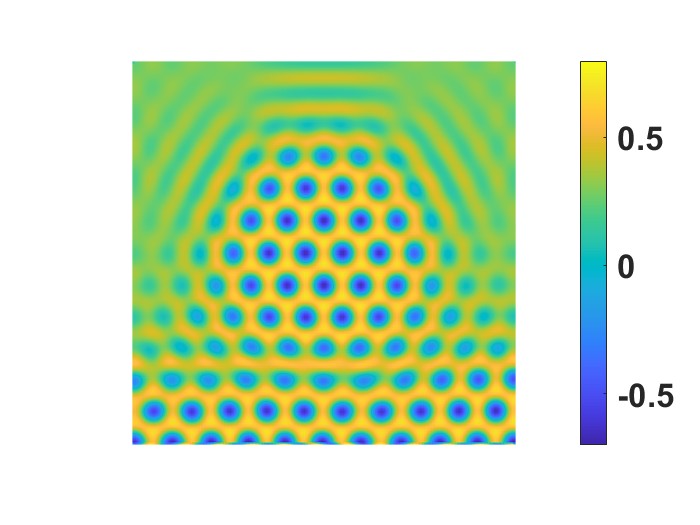}
	\end{minipage}}
	\subfigure[]{
		\begin{minipage}[b]{0.3\linewidth}
			\includegraphics[width=1\linewidth]{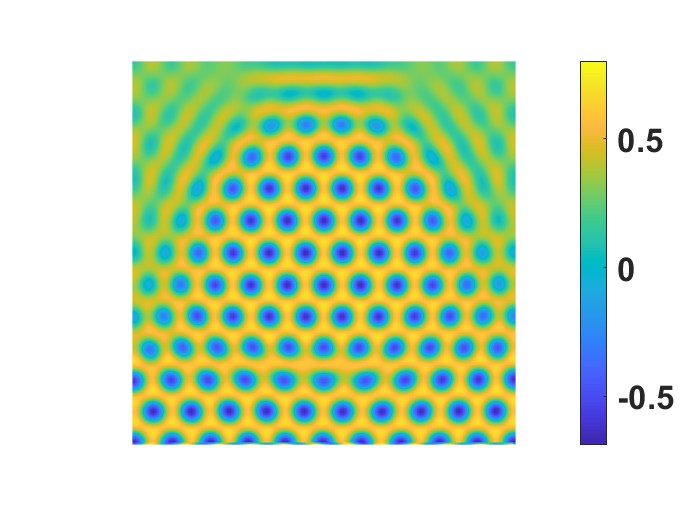}
	\end{minipage}}
	\caption{The role of model parameter $\alpha_1$ in the ordered growth. The snapshots of   $\phi$ are taken with $128 \times 128$ space meshes at $T=0, 32, 40$, respectively. Time step   $\delta t=1 \times 10^{-2}$ is used. (a)-(c): $\alpha_1=1\times 10^3,\beta_1=1$;(d)-(f): $\alpha_1=5\times 10^{-3},\beta_1=1$. A large $\alpha_1$ tends to annihilate the boundary effect to the bulk while a small $\alpha_1$ promotes crystal growth near the boundary in addition to the growth in the middle and thereby facilitates the overall growth in the domain. }\label{figure2}
\end{figure*}

We then  check the roles of $\alpha_1, \beta_1$ by benchmarking against the result in  Figure \ref{figure1}-(a-c). In Figure \ref{figure2}, a large $\alpha_1$ suppresses the roles that the surface energy and bulk energy play in inducing the across boundary mass flux at the boundary and forces a nearly homogeneous Neumann boundary condition asymptotically for $\tilde \mu$. As the result,  the surface can have very little impact on the bulk structure as shown in (a) to (c). For a small $\alpha_1$, on the other hand, the difference between the surface energy and bulk energy is amplified at the boundary to lead to a large across boundary  mass flux. As the result, the crystal growth at the boundary is significantly accelerated as shown in Figure \ref{figure2}-(d) to (f).

\begin{figure*}
	\centering
	\subfigure[]{
		\begin{minipage}[b]{0.3\linewidth}
			\includegraphics[width=1\linewidth]{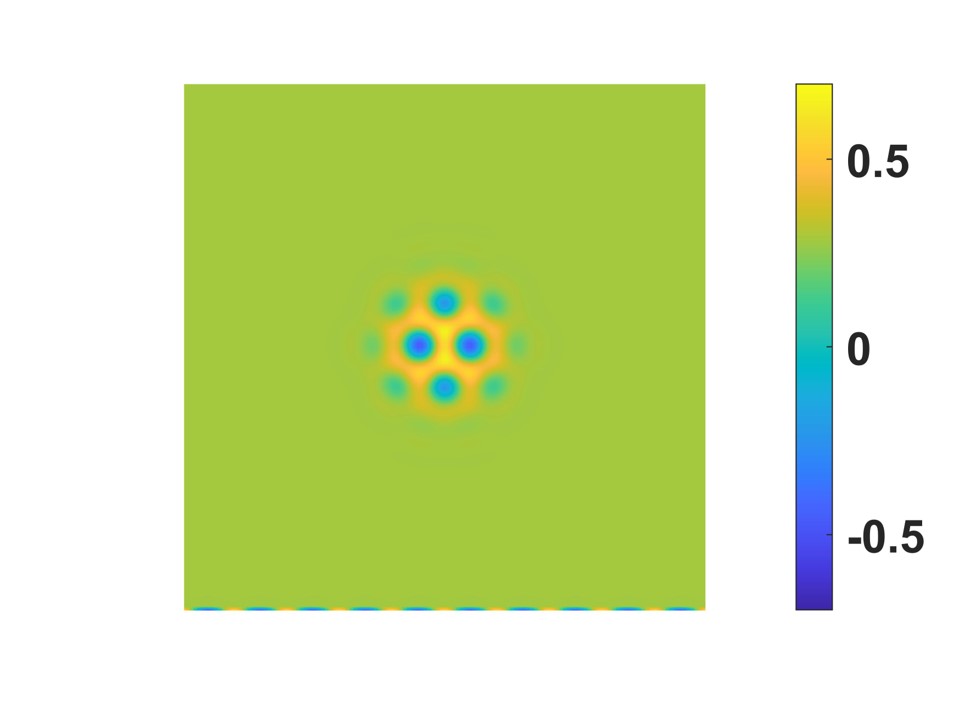}
	\end{minipage}}
	\subfigure[]{
		\begin{minipage}[b]{0.3\linewidth}
			\includegraphics[width=1\linewidth]{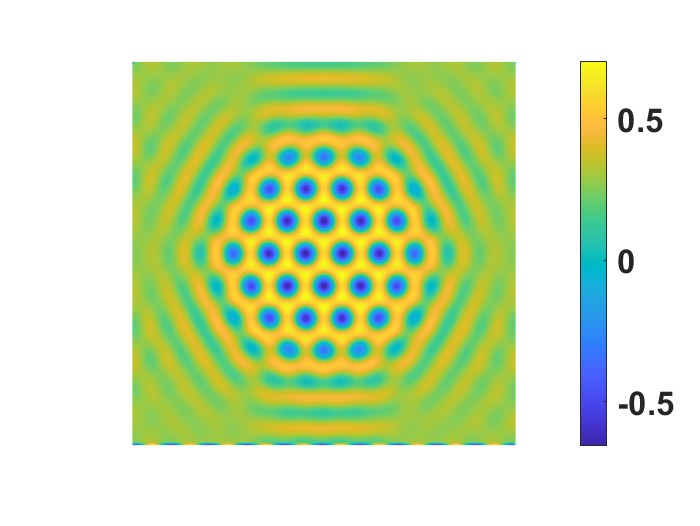}
	\end{minipage}}
	\subfigure[]{
		\begin{minipage}[b]{0.3\linewidth}
			\includegraphics[width=1\linewidth]{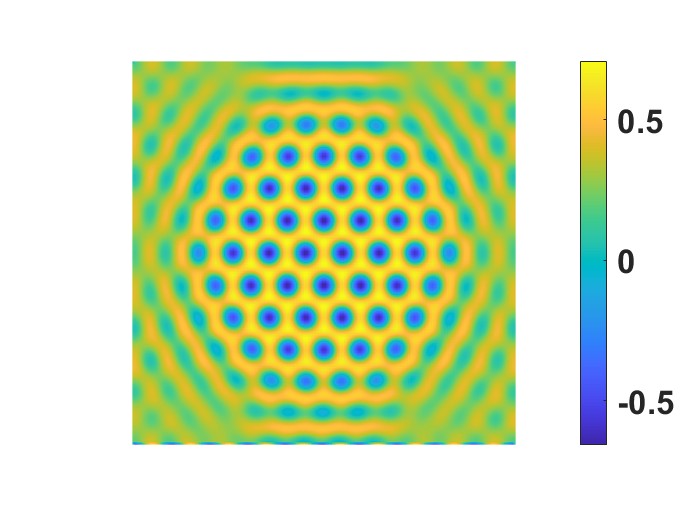}
	\end{minipage}}
	\subfigure[]{
		\begin{minipage}[b]{0.3\linewidth}
			\includegraphics[width=1\linewidth]{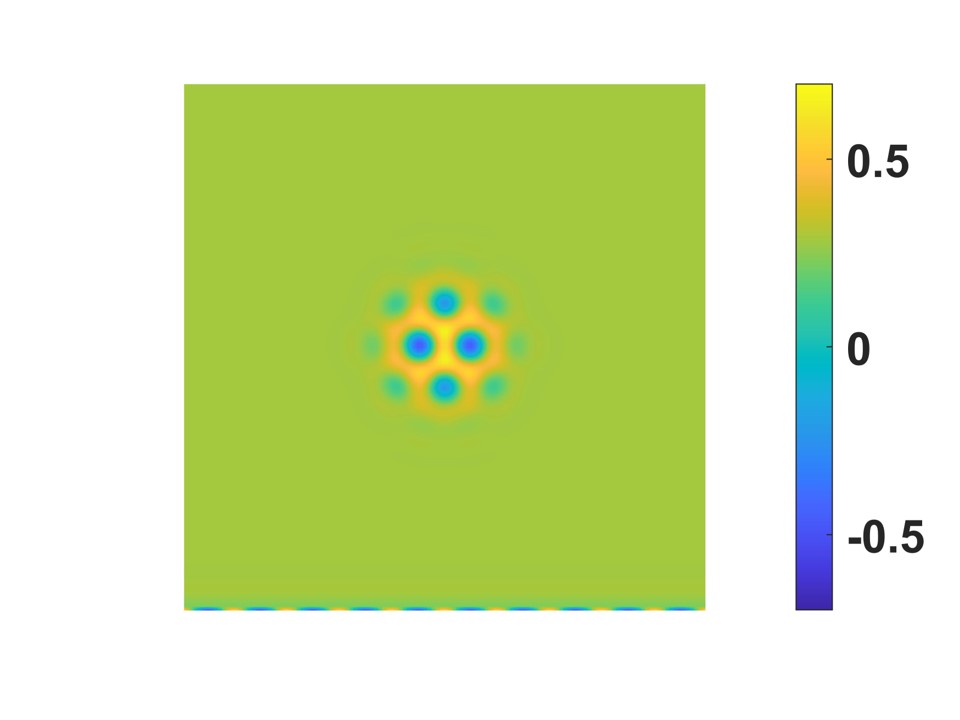}
	\end{minipage}}
	\subfigure[]{
		\begin{minipage}[b]{0.3\linewidth}
			\includegraphics[width=1\linewidth]{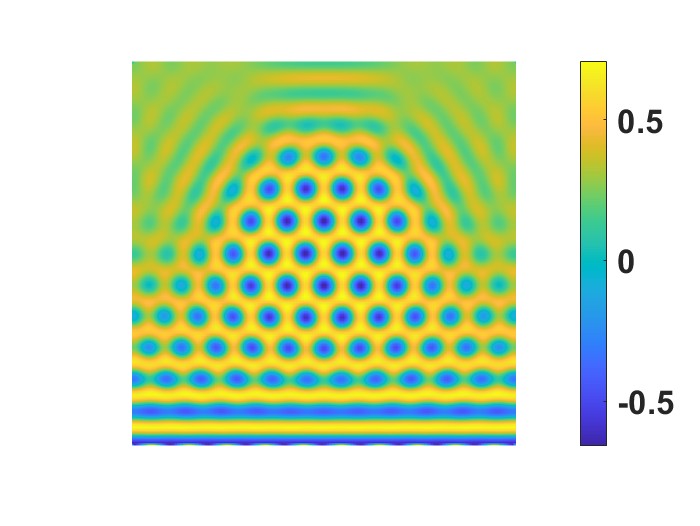}
	\end{minipage}}
	\subfigure[]{
		\begin{minipage}[b]{0.3\linewidth}
			\includegraphics[width=1\linewidth]{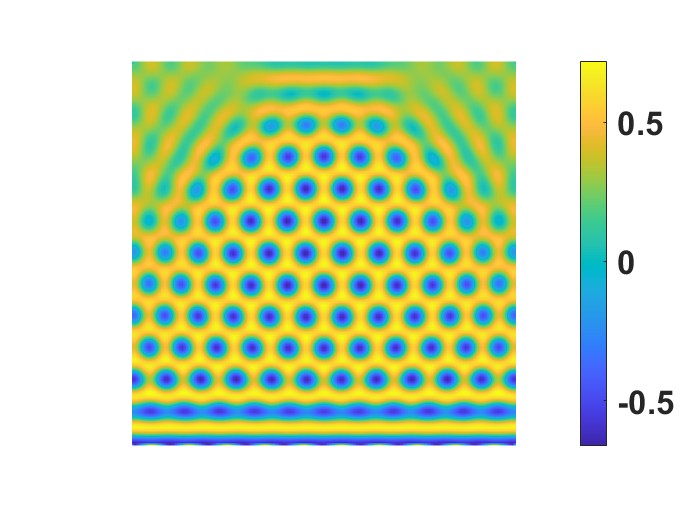}
	\end{minipage}}
	\caption{The role of parameter $\beta_1$ in the ordered growth. Snapshots of $\phi$ are taken  with $128 \times 128$ space meshes at $T=0, 32, 40$, respectively. Time step is chosen as $\delta t=1 \times 10^{-2}$. (a)-(c):   $\alpha_1=1,\beta_1=10$; (d)-(f):    $\alpha_1=1,\beta_1=1\times 10^{-4}$. A large $\beta_1$ diminishes the effect of boundary dynamics while a small $\beta_1$ facilitates growth near the boundary in a different pattern.  }\label{figure3}
\end{figure*}

\begin{figure*}
	\centering
	\subfigure[]{
		\begin{minipage}[b]{0.44\linewidth}
			\includegraphics[width=1\linewidth]{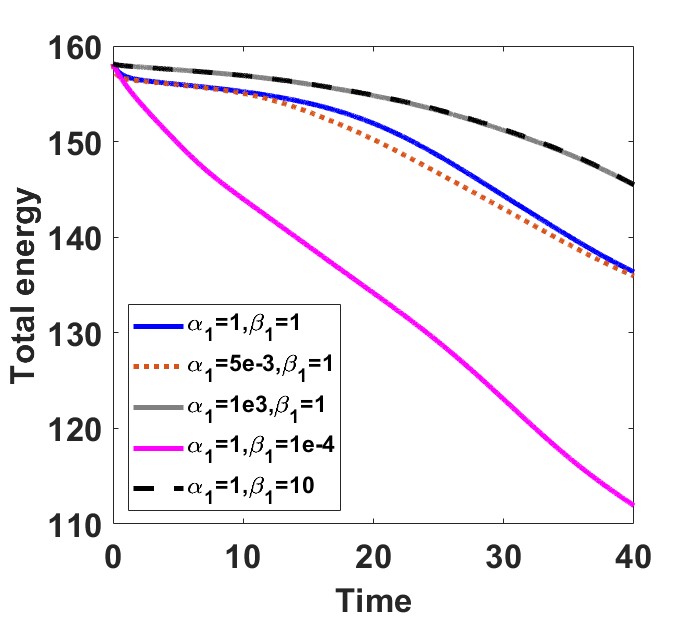}
	\end{minipage}}
	\subfigure[]{
		\begin{minipage}[b]{0.44\linewidth}
			\includegraphics[width=1\linewidth]{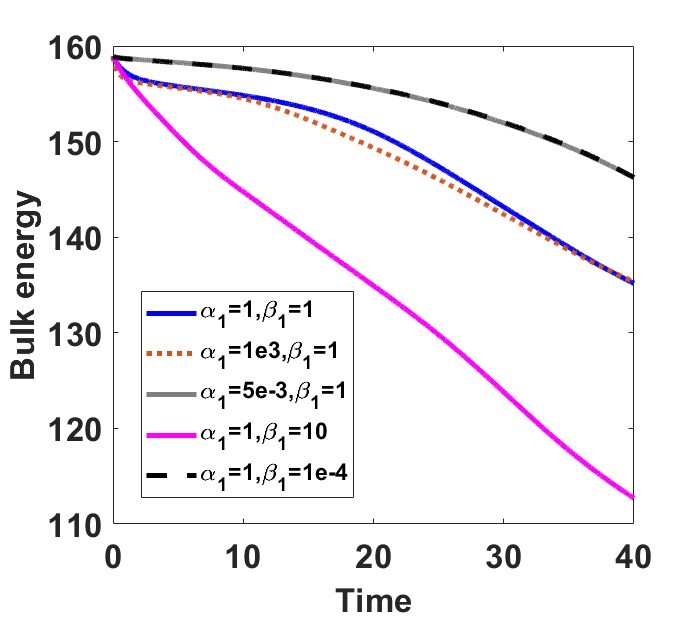}
	\end{minipage}}
	\subfigure[]{
		\begin{minipage}[b]{0.44\linewidth}
			\includegraphics[width=1\linewidth]{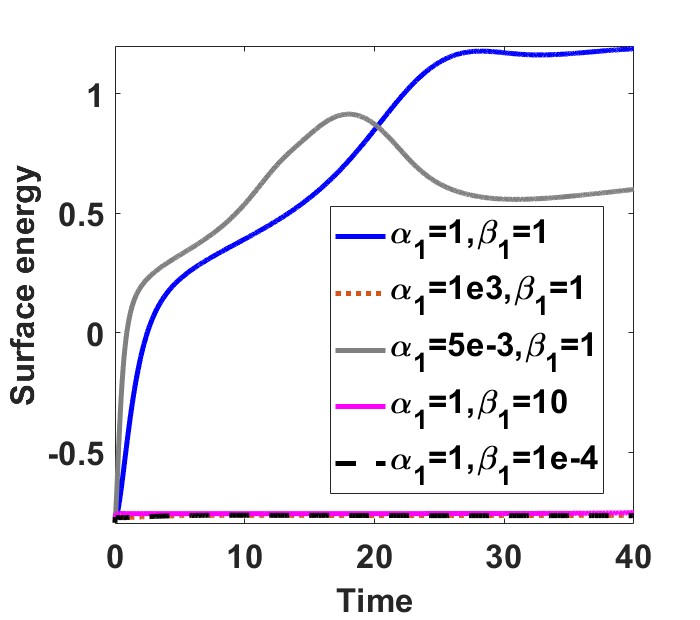}
	\end{minipage}}
	\subfigure[]{
		\begin{minipage}[b]{0.44\linewidth}
			\includegraphics[width=1\linewidth]{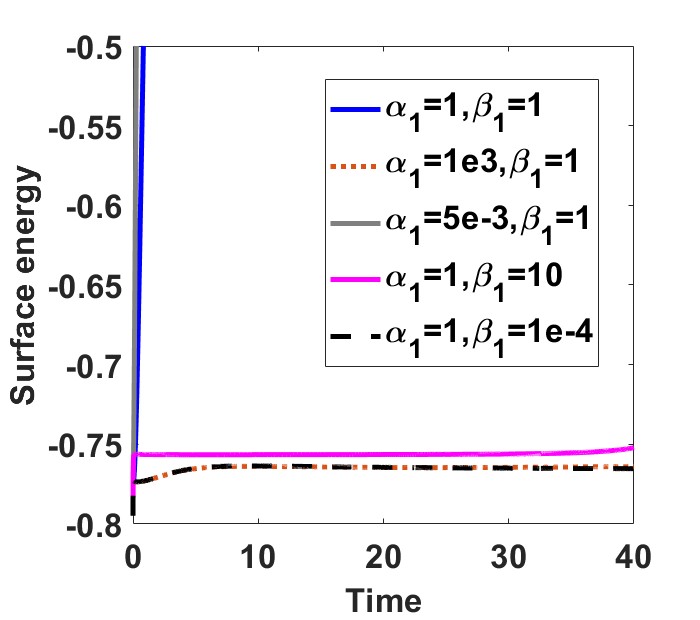}
	\end{minipage}}
	\caption{Time evolutions of total energy, bulk energy and surface energy with different $\alpha_1, \beta_1$ in (a-c). In (d), the range of surface energies are between $[-0.8,-0.5]$, which is a supplementary figure for (c). Though the surface energies for $\alpha_1=1\times 10^3, \beta_1=1$ and $\alpha_1=1, \beta_1=1\times 10^{-4}$ are almost same, their bulk free energies are different. The bulk free energies for $\alpha_1=5\times 10^{-3}, \beta_1=1$ and $\alpha_1=1, \beta_1=1\times 10^{-4}$ are also almost same, but their surface free energies are different. It reveals the magnitudes of the bulk and surface energies and the weights for chemical potentials or flux determine the patterns.}\label{figure4}
\end{figure*}

$\beta_1$ is also varied while $\alpha_1$ is fixed to show the effect of the surface energy on the bulk pattern in Figure \ref{figure3}. If $\beta_1$ is large,  it forces a near static boundary state with $\mu_s \approx 0$. That is the reason why the pattern in the bulk is similar to the one with the homogeneous Neumann boundary condition shown in Figure \ref{figure2}-(a-c). If  $\beta_1$ is small, crystal growth near the surface tends to form  solid crystallites with weak hexagonal ordering resembling a  lamellar pattern shown  in Figure \ref{figure3}-(d-f). This prevents the well-ordered crystal from growing into the boundary region. The time evolutions of total energy,  bulk energy and surface energy in figure \ref{figure4} show that the total free energy dissipation is guaranteed. However, the surface energy may increase due to the mass transfer at the boundary. The patterns in figure \ref{figure2}-(a-c) and \ref{figure3}-(a-c) are similar is because their bulk energies are the same as in \ref{figure4}-(b), however, their surface free energies are different. The difference of surface energies between them are covered by the magnitude of the bulk energy. Figure \ref{figure4} show numerically that both  $\alpha_1$ and $\beta_1$ control the magnitude of the energies. 

In the example, we demonstrate that the surface energy and prescribed dynamic boundary conditions can indeed influence bulk dynamics in various ways depending on what surface physical effects are dominating. It paves the way for one to alter or even manipulate bulk dynamics by controlling the boundary condition especially when the bulk energy and surface energy become comparable in a small confined geometry.
\clearpage

\section{Conclusions}\label{sec5}

\noindent \indent We have presented a hierarchical procedure for deriving thermodynamically consistent models together with consistent dynamic boundary conditions using the generalized Onsager principle in tandem. We illustrate it using the binary phase field model with up to second order spatial derivatives in the free energy functional. Extensions to  models with more general free energy functionals, including the one with nonlocal interaction kernels, can be derived following an analogous procedure. We show that many existing binary phase field models with the dynamic/static boundary conditions are in fact special cases of the general model. We then show the effect of the surface energy and the dynamic boundary conditions on solutions in the bulk in crystal growth processes through numerical simulations. This study  summarizes a thermodynamically consistent  modeling protocol. It also paves the way for one to develop structural preserving, thermodynamically consistent numerical algorithms for the resulting models.

\section*{Acknowledgements}
Xiaobo Jing's research is partially supported by
NSFC awards \#11971051, \#12147165 and NSAF-U1930402 to CSRC.
Qi Wang's research is partially supported by NSF awards  OIA-1655740 and a GEAR award from SC EPSCoR/IDeA Program.
\bibliographystyle{plain}
\bibliography{reference}

\begin{thebibliography}{10}

\bibitem{allen1979microscopic}
Samuel~M Allen and John~W Cahn.
\newblock A microscopic theory for antiphase boundary motion and its
  application to antiphase domain coarsening.
\newblock {\em Acta metallurgica}, 27(6):1085--1095, 1979.

\bibitem{brenner2013interfacial}
Howard Brenner.
\newblock {\em Interfacial transport processes and rheology}.
\newblock Elsevier, 2013.

\bibitem{cahn1958free}
John~W Cahn and John~E Hilliard.
\newblock Free energy of a nonuniform system. i. interfacial free energy.
\newblock {\em The Journal of chemical physics}, 28(2):258--267, 1958.

\bibitem{cavaterra2014non}
Cecilia Cavaterra, Maurizio Grasselli, and Hao Wu.
\newblock Non-isothermal viscous cahn-hilliard equation with inertial term and
  dynamic boundary conditions.
\newblock {\em Communications on Pure \& Applied Analysis}, 13(5):1855, 2014.

\bibitem{chen2017large}
Chun-Wei Chen, Chien-Tsung Hou, Cheng-Chang Li, Hung-Chang Jau, Chun-Ta Wang,
  Ching-Lang Hong, Duan-Yi Guo, Cheng-Yu Wang, Sheng-Ping Chiang, Timothy~J
  Bunning, et~al.
\newblock Large three-dimensional photonic crystals based on monocrystalline
  liquid crystal blue phases.
\newblock {\em Nature communications}, 8(1):1--8, 2017.

\bibitem{chen2002phase}
Long-Qing Chen.
\newblock Phase-field models for microstructure evolution.
\newblock {\em Annual review of materials research}, 32(1):113--140, 2002.

\bibitem{cherfils2011cahn}
Laurence Cherfils, Alain Miranville, and Sergey Zelik.
\newblock The cahn-hilliard equation with logarithmic potentials.
\newblock {\em Milan Journal of Mathematics}, 79(2):561--596, 2011.

\bibitem{colli2015cahn}
Pierluigi Colli and Takeshi Fukao.
\newblock Cahn--hilliard equation with dynamic boundary conditions and mass
  constraint on the boundary.
\newblock {\em Journal of Mathematical Analysis and Applications},
  429(2):1190--1213, 2015.

\bibitem{colli2019coupled}
Pierluigi Colli, Takeshi Fukao, and Kei~Fong Lam.
\newblock On a coupled bulk--surface allen--cahn system with an affine linear
  transmission condition and its approximation by a robin boundary condition.
\newblock {\em Nonlinear Analysis}, 184:116--147, 2019.

\bibitem{dziuk2013finite}
Gerhard Dziuk and Charles~M Elliott.
\newblock Finite element methods for surface pdes.
\newblock {\em Acta Numerica}, 22:289--396, 2013.

\bibitem{elder2007phase}
Ken~R Elder, Nikolas Provatas, Joel Berry, Peter Stefanovic, and Martin Grant.
\newblock Phase-field crystal modeling and classical density functional theory
  of freezing.
\newblock {\em Physical Review B}, 75(6):064107, 2007.

\bibitem{elder2004modeling}
KR~Elder and Martin Grant.
\newblock Modeling elastic and plastic deformations in nonequilibrium
  processing using phase field crystals.
\newblock {\em Physical Review E}, 70(5):051605, 2004.

\bibitem{elder2001sharp}
KR~Elder, Martin Grant, Nikolas Provatas, and JM~Kosterlitz.
\newblock Sharp interface limits of phase-field models.
\newblock {\em Physical Review E}, 64(2):021604, 2001.

\bibitem{fischer1997novel}
Hans~Peter Fischer, Philipp Maass, and Wolfgang Dieterich.
\newblock Novel surface modes in spinodal decomposition.
\newblock {\em Physical review letters}, 79(5):893, 1997.

\bibitem{fischer1998time}
HP~Fischer, J~Reinhard, W~Dieterich, J-F Gouyet, P~Maass, A~Majhofer, and
  D~Reinel.
\newblock Time-dependent density functional theory and the kinetics of lattice
  gas systems in contact with a wall.
\newblock {\em The Journal of chemical physics}, 108(7):3028--3037, 1998.

\bibitem{fukao2019separation}
Takeshi Fukao and Hao Wu.
\newblock Separation property and convergence to equilibrium for the equation
  and dynamic boundary condition of cahn-hilliard type with singular potential.
\newblock {\em arXiv preprint arXiv:1910.14177}, 2019.

\bibitem{fukao2017structure}
Takeshi Fukao, Shuji Yoshikawa, and Saori Wada.
\newblock Structure-preserving finite difference schemes for the cahn-hilliard
  equation with dynamic boundary conditions in the one-dimensional case.
\newblock {\em Communications on Pure \& Applied Analysis}, 16(5):1915, 2017.

\bibitem{gal2006cahn}
Ciprian~G Gal.
\newblock A cahn--hilliard model in bounded domains with permeable walls.
\newblock {\em Mathematical methods in the applied sciences},
  29(17):2009--2036, 2006.

\bibitem{gal2018doubly}
Ciprian~G Gal.
\newblock Doubly nonlocal cahn--hilliard equations.
\newblock In {\em Annales de l'Institut Henri Poincar{\'e} C, Analyse non
  lin{\'e}aire}, volume~35, pages 357--392. Elsevier, 2018.

\bibitem{gal2008asymptotic}
Ciprian~G Gal and Hao Wu.
\newblock Asymptotic behavior of a cahn-hilliard equation with wentzell
  boundary conditions and mass conservation.
\newblock {\em Discrete \& Continuous Dynamical Systems-A}, 22(4):1041, 2008.

\bibitem{galenko2005diffuse}
Peter Galenko and David Jou.
\newblock Diffuse-interface model for rapid phase transformations in
  nonequilibrium systems.
\newblock {\em Physical Review E}, 71(4):046125, 2005.

\bibitem{galenko2019rapid}
PK~Galenko and D~Jou.
\newblock Rapid solidification as non-ergodic phenomenon.
\newblock {\em Physics Reports}, 818:1--70, 2019.

\bibitem{garcke2020weak}
Harald Garcke and Patrik Knopf.
\newblock Weak solutions of the cahn--hilliard system with dynamic boundary
  conditions: A gradient flow approach.
\newblock {\em SIAM Journal on Mathematical Analysis}, 52(1):340--369, 2020.

\bibitem{gavrilyuk2019dynamic}
Sergey Gavrilyuk and Henri Gouin.
\newblock Dynamic boundary conditions for membranes whose surface energy
  depends on the mean and gaussian curvatures.
\newblock {\em Mathematics and Mechanics of Complex Systems}, 7(2):131--157,
  2019.

\bibitem{giacomin1997phase}
Giambattista Giacomin and Joel~L Lebowitz.
\newblock Phase segregation dynamics in particle systems with long range
  interactions. i. macroscopic limits.
\newblock {\em Journal of statistical Physics}, 87(1-2):37--61, 1997.

\bibitem{goldstein2011cahn}
Gis{\`e}le~Ruiz Goldstein, Alain Miranville, and Giulio Schimperna.
\newblock A cahn--hilliard model in a domain with non-permeable walls.
\newblock {\em Physica D: Nonlinear Phenomena}, 240(8):754--766, 2011.

\bibitem{grigoryan2009heat}
Alexander Grigoryan.
\newblock {\em Heat kernel and analysis on manifolds}, volume~47.
\newblock American Mathematical Soc., 2009.

\bibitem{jing2019second}
Xiaobo Jing, Jun Li, Xueping Zhao, and Qi~Wang.
\newblock Second order linear energy stable schemes for allen-cahn equations
  with nonlocal constraints.
\newblock {\em Journal of Scientific Computing}, 80(1):500--537, 2019.

\bibitem{jing2020linear}
Xiaobo Jing and Qi~Wang.
\newblock Linear second order energy stable schemes for phase field crystal
  growth models with nonlocal constraints.
\newblock {\em Computers \& Mathematics with Applications}, 79(3):764--788,
  2020.

\bibitem{jing2022DBC}
Xiaobo Jing and Qi~Wang.
\newblock Structure preserving algorithms for thermodynamically consistent
  partial differential equation with dynamical boundary conditions.
\newblock {\em to be submitted}, 2022.

\bibitem{karma1999phase}
Alain Karma and Wouter-Jan Rappel.
\newblock Phase-field model of dendritic sidebranching with thermal noise.
\newblock {\em Physical review E}, 60(4):3614, 1999.

\bibitem{kenzler2001phase}
Rainer Kenzler, Frank Eurich, Philipp Maass, Bernd Rinn, Johannes Schropp,
  Erich Bohl, and Wolfgang Dieterich.
\newblock Phase separation in confined geometries: Solving the cahn--hilliard
  equation with generic boundary conditions.
\newblock {\em Computer Physics Communications}, 133(2-3):139--157, 2001.

\bibitem{kim2012phase}
Junseok Kim.
\newblock Phase-field models for multi-component fluid flows.
\newblock {\em Communications in Computational Physics}, 12(3):613--661, 2012.

\bibitem{knopf2019convergence}
Patrik Knopf and Kei~Fong Lam.
\newblock Convergence of a robin boundary approximation for a cahn--hilliard
  system with dynamic boundary conditions.
\newblock {\em arXiv preprint arXiv:1908.06124}, 2019.

\bibitem{knopf2020phase}
Patrik Knopf, Kei~Fong Lam, Chun Liu, and Stefan Metzger.
\newblock Phase-field dynamics with transfer of materials: The cahn--hillard
  equation with reaction rate dependent dynamic boundary conditions.
\newblock {\em arXiv preprint arXiv:2003.12983}, 2020.

\bibitem{knopf2021nonlocal}
Patrik Knopf and Andrea Signori.
\newblock On the nonlocal cahn--hilliard equation with nonlocal dynamic
  boundary condition and boundary penalization.
\newblock {\em Journal of Differential Equations}, 280:236--291, 2021.

\bibitem{kundin2017application}
Julia Kundin and Muhammad~Ajmal Choudhary.
\newblock Application of the anisotropic phase-field crystal model to
  investigate the lattice systems of different anisotropic parameters and
  orientations.
\newblock {\em Modelling and Simulation in Materials Science and Engineering},
  25(5):055004, 2017.

\bibitem{li2003thin}
Bo~Li and Jian-Guo Liu.
\newblock Thin film epitaxy with or without slope selection.
\newblock {\em European Journal of Applied Mathematics}, 14(6):713--743, 2003.

\bibitem{li2019energy}
Jun Li, Jia Zhao, and Qi~Wang.
\newblock Energy and entropy preserving numerical approximations of
  thermodynamically consistent crystal growth models.
\newblock {\em Journal of Computational Physics}, 382:202--220, 2019.

\bibitem{liu2019energetic}
Chun Liu and Hao Wu.
\newblock An energetic variational approach for the cahn--hilliard equation
  with dynamic boundary condition: model derivation and mathematical analysis.
\newblock {\em Archive for Rational Mechanics and Analysis}, 233(1):167--247,
  2019.

\bibitem{martinez2015blue}
Jos{\'e}~A Mart{\'\i}nez-Gonz{\'a}lez, Ye~Zhou, Mohammad Rahimi, Emre
  Bukusoglu, Nicholas~L Abbott, and Juan~J de~Pablo.
\newblock Blue-phase liquid crystal droplets.
\newblock {\em Proceedings of the National Academy of Sciences},
  112(43):13195--13200, 2015.

\bibitem{nestler2011phase}
Britta Nestler and Abhik Choudhury.
\newblock Phase-field modeling of multi-component systems.
\newblock {\em Current opinion in solid state and Materials Science},
  15(3):93--105, 2011.

\bibitem{okumura2020structure}
Makoto Okumura and Daisuke Furihata.
\newblock A structure-preserving scheme for the allen--cahn equation with a
  dynamic boundary condition.
\newblock {\em Discrete \& Continuous Dynamical Systems-A}, 40(8):4927, 2020.

\bibitem{onsager1931reciprocal1}
Lars Onsager.
\newblock Reciprocal relations in irreversible processes. i.
\newblock {\em Physical review}, 37(4):405, 1931.

\bibitem{onsager1931reciprocal2}
Lars Onsager.
\newblock Reciprocal relations in irreversible processes. ii.
\newblock {\em Physical review}, 38(12):2265, 1931.

\bibitem{onsager1953fluctuations}
Lars Onsager and Stefan Machlup.
\newblock Fluctuations and irreversible processes.
\newblock {\em Physical Review}, 91(6):1505, 1953.

\bibitem{provatas2011phase}
Nikolas Provatas and Ken Elder.
\newblock {\em Phase-field methods in materials science and engineering}.
\newblock John Wiley \& Sons, 2011.

\bibitem{rahman2015blue}
MD~Asiqur Rahman, Suhana~Mohd Said, and S~Balamurugan.
\newblock Blue phase liquid crystal: strategies for phase stabilization and
  device development.
\newblock {\em Science and technology of advanced materials}, 16(3):033501,
  2015.

\bibitem{rakita2019defects}
Yevgeny Rakita, Igor Lubomirsky, and David Cahen.
\newblock When defects become dynamic: halide perovskites: a new window on
  materials?
\newblock {\em Materials Horizons}, 6(7):1297--1305, 2019.

\bibitem{rybka1999convergence}
Piotr Rybka and Karl-Heinz Hoffnlann.
\newblock Convergence of solutions to cahn-hilliard equation.
\newblock {\em Communications in partial differential equations},
  24(5-6):1055--1077, 1999.

\bibitem{salhoumi2016gibbs}
A~Salhoumi and PK~Galenko.
\newblock Gibbs--thomson condition for the rapidly moving interface in a binary
  system.
\newblock {\em Physica A: Statistical Mechanics and its Applications},
  447:161--171, 2016.

\bibitem{singer2008phase}
Irina Singer-Loginova and HM~Singer.
\newblock The phase field technique for modeling multiphase materials.
\newblock {\em Reports on progress in physics}, 71(10):106501, 2008.

\bibitem{steinbach2009phase}
Ingo Steinbach.
\newblock Phase-field models in materials science.
\newblock {\em Modelling and simulation in materials science and engineering},
  17(7):073001, 2009.

\bibitem{steinbach1996phase}
Ingo Steinbach, Franco Pezzolla, Britta Nestler, Markus See{\ss}elberg, Robert
  Prieler, Georg~J Schmitz, and Joao~LL Rezende.
\newblock A phase field concept for multiphase systems.
\newblock {\em Physica D: Nonlinear Phenomena}, 94(3):135--147, 1996.

\bibitem{sun2020structure}
Shouwen Sun, Jun Li, Jia Zhao, and Qi~Wang.
\newblock Structure-preserving numerical approximations to a non-isothermal
  hydrodynamic model of binary fluid flows.
\newblock {\em Journal of Scientific Computing}, 83:50, 2020.

\bibitem{Wang2020}
Qi~Wang.
\newblock Generalized onsager principle and its application.
\newblock In Xiang you Liu, editor, {\em Frontiers and Progress of Current Soft
  Matter Research}. Springer Nature, 2020.

\bibitem{warren1995prediction}
James~A Warren and William~J Boettinger.
\newblock Prediction of dendritic growth and microsegregation patterns in a
  binary alloy using the phase-field method.
\newblock {\em Acta Metallurgica et Materialia}, 43(2):689--703, 1995.

\bibitem{wu2021review}
Hao Wu.
\newblock A review on the cahn-hilliard equation: Classical results and recent
  advances in dynamic boundary conditions.
\newblock {\em arXiv preprint arXiv:2112.13812}, 2021.

\bibitem{xing2009topology}
Xiangjun Xing.
\newblock Topology and geometry of smectic order on compact curved substrates.
\newblock {\em Journal of Statistical Physics}, 134(3):487--536, 2009.

\bibitem{yang2017numerical}
Xiaofeng Yang, Jia Zhao, Qi~Wang, and Jie Shen.
\newblock Numerical approximations for a three-component cahn--hilliard
  phase-field model based on the invariant energy quadratization method.
\newblock {\em Mathematical Models and Methods in Applied Sciences},
  27(11):1993--2030, 2017.

\bibitem{yang2016hydrodynamic}
Xiaogang Yang, Jun Li, M~Gregory Forest, and Qi~Wang.
\newblock Hydrodynamic theories for flows of active liquid crystals and the
  generalized onsager principle.
\newblock {\em Entropy}, 18(6):202, 2016.

\bibitem{yoon2011topology}
Gil~Ho Yoon.
\newblock Topology optimization for nonlinear dynamic problem with multiple
  materials and material-dependent boundary condition.
\newblock {\em Finite elements in analysis and design}, 47(7):753--763, 2011.

\bibitem{zhao2017numerical}
Jia Zhao, Qi~Wang, and Xiaofeng Yang.
\newblock Numerical approximations for a phase field dendritic crystal growth
  model based on the invariant energy quadratization approach.
\newblock {\em International Journal for Numerical Methods in Engineering},
  110(3):279--300, 2017.

\bibitem{zhao2018general}
Jia Zhao, Xiaofeng Yang, Yuezheng Gong, Xueping Zhao, Xiaogang Yang, Jun Li,
  and Qi~Wang.
\newblock A general strategy for numerical approximations of non-equilibrium
  models-part i: thermodynamical systems.
\newblock {\em Int. J. Numer. Anal. Model}, 15(6):884--918, 2018.

\bibitem{zhao2018thermodynamically}
Xueping Zhao, Tiezheng Qian, and Qi~Wang.
\newblock Thermodynamically consistent hydrodynamic models of multi-component
  fluid flows.
\newblock {\em arXiv preprint arXiv:1809.05494}, 2018.

\end{thebibliography}
\end{document}